\documentclass[11pt]{article}
\pdfoutput = 1

\usepackage[utf8]{inputenc}
\usepackage{color,graphicx}
\usepackage{verbatim}
\usepackage{enumerate}
\usepackage{amssymb}
\usepackage{stmaryrd}
\usepackage{physics}
\usepackage{amsfonts}
\usepackage{cite}
\usepackage{array}
\usepackage{setspace}
\usepackage{float}
\usepackage{url}
\usepackage{mathtools}
\usepackage{bbold}
\usepackage{tikz}
\usepackage{graphicx}
\usepackage{wrapfig}
\usepackage{subfigure}
\usepackage[labelfont=bf,font={rm},font={small}]{caption}

\usepackage{tikz}
\usepackage{cite}
\allowdisplaybreaks

\usepackage[margin = 2.2cm]{geometry}
\usepackage[ragged]{footmisc}
\usepackage{hyperref}
\hypersetup{
    colorlinks,
    citecolor=blue,
    filecolor=black,
    linkcolor=blue,
    urlcolor=blue,
    linktocpage=true
}

\def\l{\lambda}
\def\G{\Gamma}

\newcommand{\be}{\begin{equation}}
\newcommand{\ee}{\end{equation}}
\newcommand{\bea}{\begin{align}}
\newcommand{\eea}{\end{align}}
\newcommand{\bi}{\begin{itemize}}
\newcommand{\ei}{\end{itemize}}

\newcommand{\lr}[1]{\left( #1 \right)}

\def\t{\text}

\def\rb{\rangle}
\def\lb{\langle}
\def\eschw{E_{\t{schw}}}

\def\Xint#1{\mathchoice
   {\XXint\displaystyle\textstyle{#1}}%
   {\XXint\textstyle\scriptstyle{#1}}%
   {\XXint\scriptstyle\scriptscriptstyle{#1}}%
   {\XXint\scriptscriptstyle\scriptscriptstyle{#1}}%
   \!\int}
\def\XXint#1#2#3{{\setbox0=\hbox{$#1{#2#3}{\int}$}
     \vcenter{\hbox{$#2#3$}}\kern-.5\wd0}}
\def\dashint{\Xint-}

\numberwithin{equation}{section}

\title{Template}

\begin{document}

\thispagestyle{empty}
\begin{center}
\vspace*{.4cm}
  

    {\LARGE \bf 
God of the Gaps:\bigskip\\Random matrix models and the black hole spectral gap
  }
    
    \vspace{0.4in}
    {\bf Clifford V. Johnson$^{\dagger}$, Mykhaylo Usatyuk$^{*}$}

\bigskip\bigskip
    
{$^{\dagger}$Physics Department, Broida Hall, University of California, Santa Barbara, CA 93106, USA}
    \vspace{0.1in}
    
{$^{*}$Kavli Institute for Theoretical Physics, Kohn Hall, Santa Barbara, CA 93106, USA}
    \vspace{0.1in}

    {\tt cliffordjohnson@ucsb.edu},  {\tt musatyuk@kitp.ucsb.edu}
\end{center}

\vspace{0.4in}
\begin{abstract}
\noindent We show that random matrix models are a natural tool for understanding the appearance  of a large gap in the microstate spectrum of black holes  when there is a high degeneracy of  states, in a variety of settings. While the most natural context is extended supersymmetry, where the number of BPS states scales  as ${\rm e}^{S_0}$, where  $S_0$  is the $T{=}0$ entropy, it is a robust feature that a large gap will appear whenever there is a mechanism producing a  high degree of degeneracy.  In random matrix model terms, the phenomenon is simply an extreme case of eigenvalue repulsion in the effective log gas description. We exhibit several examples for illustration, starting with the simple  Wishart model, continuing with extensions of it that incorporate multicritical behaviour  allowing for the emergence of gravity, and culminating in   constructing multicritical matrix models of ${\cal N}{=}2$ and  ${\cal N}{=}4$ JT supergravity theories, the latter of which is new.
\end{abstract}
\pagebreak
\setcounter{page}{1}
\tableofcontents

\newpage
\section{Introduction}

\label{sec:introduction}

\subsection{The fate of extremal black holes} 

Recent exciting progress~\cite{Iliesiu:2020qvm,Heydeman:2020hhw,Boruch:2022tno} in understanding the low temperature thermodynamics of extremal black holes  in semi-classical quantum gravity has highlighted the key role of the Schwarzian mode that arises in  the throat description of the near-horizon geometry.\footnote{For an excellent brief summary, see ref.~\cite{Turiaci:2023wrh}.} The Schwarzian dynamics can be  exactly solved quantum mechanically~\cite{Cotler:2016fpe,Maldacena:2016hyu,Maldacena:2016upp,Mertens:2017mtv,Stanford:2017thb}, and the resulting one-loop correction to the leading physics significantly modifies the low temperature dynamics of the black hole, successfully addressing two age-old concerns about the physics of such black holes: 

{\bf (1)} The naive gravitational path integral (GPI) computation~\cite{Gibbons:1976ue}, through the entropy formula of Bekenstein and Hawking~\cite{Bekenstein:1973ur,Hawking:1976de},  $S{=}A/4G_N$,  suggests that since the area $A$ of the black hole is non-zero at $T=0$  there is a non-zero entropy there, (denoted $S_0$ henceforth). Interpreted as a stunningly large ground state degeneracy, such a quantity is hard to understand (see {\it e.g.,} ref.~\cite{Page:2000dk}) without some additional properties of the theory such as a symmetry that allows or protects it. Instead, incorporating the Schwarzian mode into the GPI analysis~\cite{Iliesiu:2020qvm} drastically modifies the density of states below a certain temperature/energy scale where the naive analysis breaks down, and it goes to zero at $E=0$, strongly suggesting that extremal black holes are generically absent from the theory.\footnote{This is consistent with other recent results~\cite{Horowitz:2022mly,Horowitz:2023xyl,Horowitz:2024dch,Horowitz:2022leb} that suggest that near-extremal black holes are very sensitive to developing singularities on their horizons due to the addition of higher curvature corrections or generic asymmetric perturbations.}

{\bf (2)} The need for new physics of some kind, appearing below some ``breakdown'' temperature scale, was argued for a long ago in ref.~\cite{Preskill:1991tb}, on the grounds that without it, during the late stages of evaporation the extremal black hole would be able to emit quanta carrying a fraction of the available energy that is larger than permitted by the assumption that the overall black hole system is a thermal object. So removing extremal black holes altogether certainly removes that potential paradox.  However, there {\it are} situations where extremal black holes exist in gravity, and in fact they are one of the pillars of success of (super)string theory as a quantum theory of gravity: As {\it BPS states} in supersymmetric gravity,  they allow for a precise computation of their entropy $S_0$ using counting arguments~\cite{Strominger:1996sh}: Simply put,~${\rm e}^{S_0}$ is the count of the number of ways of realizing such an extremal black hole as a threshold bound state of D-branes (along with some internal momentum, as well as  KK charge or NS-charge for the $D{=}4$ prototype~\cite{Johnson:1996ga,Maldacena:1996gb}), themselves BPS states of the underlying quantum gravity description~\cite{Polchinski:1995mt}.  Since extremal black holes appear to be necessary in such scenarios with extended supersymmetry, the concerns of ref.~\cite{Preskill:1991tb} cannot be avoided by simply removing them from the board. 

Instead something more subtle must happen, as first emphasised in ref.~\cite{Maldacena:1996ds,Maldacena:1998uz}: Crucially there must be a {\it gap in the spectrum}, immediately above the (degenerate) ground state at zero energy, with black hole excitation states only appearing above the breakdown scale. The details of how the gap scale is realized depends upon the details of how the black hole is embedded in the string theory (or, perhaps, other quantum gravity scenarios), but the need for it is clear.

Remarkably, the presence of a gap was confirmed using semiclassical techniques  for (near) BPS black holes in four dimensional ${\cal N}{=}2$ supergravity in ref.~\cite{Heydeman:2020hhw}  by including the appropriate (${\cal N}{=}4$  supersymmetric)  Schwarzian dynamics into the GPI analysis. The density of states develops a gap in the spectrum of precisely the kind needed.  That initial example was followed by the results of  ref.~\cite{Boruch:2022tno} for  (near) 1/16-BPS states of ${\cal N}{=}4$ super Yang Mills, dual to near extreme black holes in AdS$_5$, where in that case the  physics involves the ${\cal N}{=}2$ supersymmetric  Schwarzian. 

In each case, there is  a large ground state degeneracy, proportional to ${\rm e}^{S_0}$ where $S_0$ is the black hole entropy.  The key observation of this paper is that the two phenomena (large degeneracy and gravity spectral gap) go {\it hand in hand}. The core chain of reasoning  is that the effective low energy near-horizon dynamics has a random matrix model description, and certain robust  features of random matrix models of Hamiltonians with a large number of ground states {\it ensures} that in the semi-classical limit, a gap must appear as a direct result of the degeneracy. \textbf{Note added:} See ref.~\cite{Hernandez-Cuenca:2024icn} for a related recent discussion on the emergence of the gap in the black hole spectrum due to a large BPS degeneracy.

\subsection{The role of random matrix models} 

In the above black hole examples, the effective reduced descriptions of the AdS$_2{\times}S^n$ ($n{=}2,3$) throat dynamics of the black holes' near-horizon low temperature physics  are   supersymmetric extensions of Jackiw-Teitelboim gravity~\cite{Jackiw:1984je,Teitelboim:1983ux} (JT supergravity short) a two dimensional theory consisting of coupling a scalar $\Phi$ (representing the leading dynamics of the overall  size of the sphere~$S^n$)  to the 2D AdS$_2$ metric in such a way as to induce on the boundary the  Schwarzian action when $\Phi$ is integrated out.

On the other hand, there has been growing understanding that models of $D{=}2$ dilaton gravity in this class can be described perturbatively (in topology) and beyond as special classes of random matrix models. This began with the case of ordinary JT gravity and ${\cal N}{=}1$ JT supergravity  in ref.~\cite{Saad:2019lba} and ref.~\cite{Stanford:2019vob} respectively. More recently ref.~\cite{Turiaci:2023jfa} presented detailed arguments that ${\cal N}{=}2$ JT has a random matrix model description (at least perturbatively in a genus expansion) and presented some preliminary evidence that this may also be true for the ${\cal N}{=}4$ case. 

Crucially, the supersymmetric models were shown (in refs.~\cite{Johnson:2020heh,Johnson:2020exp} for ${\cal N}{=}1$ and more recently ref.~\cite{Johnson:2023ofr} for ${\cal N}{=}2$) to have an extremely natural {\it fully non-perturbative} description in terms of being built from certain multicritical models that were explored long ago in refs.~\cite{Morris:1991cq,Morris:1990bw,Dalley:1992qg,Dalley:1991vr}. 
Central to that work is the fact that a non-linear differential equation of a very special form admits just the right kinds of solutions appropriate to  each case. (This paper will show that this is also true in the case of ${\cal N}{=}4$ JT supergravity, yielding a complete non-perturbative matrix model description of it.) In ref.~\cite{Johnson:2021rsh}, while studying properties of the ``Bessel models'', special cases of the models of this class\footnote{In a sense all the models in this class are founded on the form of the  basic Wishart model\cite{10.2307/2331939}, as will be discussed below.} it became clear that they could be natural toy models of the gap seen in black hole physics, since  there is a gap in the spectrum that grows with the order, $\Gamma$, of the Bessel function used to build the spectral density. Integer~$\Gamma$ is   a count of the ground states in that model, which is suggestive.  However, it  emerged that the  non-trivial manner~\cite{Johnson:2023ofr} in which this class of models  can be used to construct the precise spectrum of the ${\cal N}{=}2$ (and as will be seen in this paper, ${\cal N}{=}4$)
cases suggests a stronger, more compelling statement, which is the core message of this paper: Random matrix models {\it predict} that there must be a large gap in the supergravity black hole spectrum when there is a large degeneracy, ${\cal O}({\rm e}^{S_0})$,  of ground states. 

The key mechanism is as follows: The effective Dyson gas description of the energies in the model has a logarithmic repulsion between any two energies generically, a famous phenomenon responsible for the characteristic (chaotic) random matrix model spectrum. However, for matrices of the right class, where there is some built-in degeneracy at some energy, there is an enhanced logarithmic repulsion force pushing all other energies away from the degenerate point.

More specifically, in modeling gravity with ${\cal O}({\rm e}^{S_0})$ degrees of freedom, the matrix model is made from  $N\times N$ matrices, where $N\sim {\rm e}^{S_0}$ is a large number. Hence, the generic microstate separation is $\sim {\rm e}^{-S_0}$, enforced by repulsion, but invisible in semi-classical physics. The spectral result at leading order is a smooth droplet solution for the Dyson gas in the leading large~$N$ limit, {\it i.e.,} semi-classical gravity. Imagine that a significant number, $\Gamma$, of eigenvalues  are degenerate such that $\Gamma/N$ is finite at large~$N$, {\it i.e.} $\Gamma$ is ${\cal O}({\rm e}^{S_0})$. In such a case, due to the repulsion mechanism,  the natural gap size in the vicinity of  this degeneracy is  amplified by at least that factor, ceasing therefore to be microscopic, becoming visible in the semi-classical gravity regime, and   should  inevitably play an important dynamical role. This is precisely what happens for BPS extremal black holes, and indeed {\it must} happen according to the supergravity considerations reviewed in the previous subsection. Here we see that since the low lying spectrum is controlled by the effective 2D gravity theory, and that this can be expected to be captured by a random matrix model, it follows that the gap's presence is robust from the matrix model perspective.

Notice that the repulsion mechanism, while natural in the supersymmetric setting, is not really strongly tied to its presence. This suggests that  it could be relevant to possible non-supersymmetric gravitational situations where a large degeneracy is protected by (perhaps) another mechanism. Especially if the low-lying dynamics results in an effective low dimensional gravity description, the random matrix model prediction would be that a spectral gap {\it must} appear. 

\subsection{Outline}

After reviewing some generalities about the form of the measure for integration over eigenvalues  in section~\ref{sec:measure4measure},  we discuss in section~\ref{sec:wishart-and-more} the most simple prototype, the Wishart model, along with some generalizations that will be useful later.   We discuss in section~\ref{sec:dyson-gas} how the Dyson gas picture guides intuition about how a large spectral gap can develop in the classical limit, through eigenvalue repulsion.  We then, in section~\ref{sec:orthog-poly}, review and develop a more powerful set of methods (orthogonal polynomial on the half line) that will serve us well later when discussing gravity, and also when formulating all the physics non-perturbatively. In that language, gapped spectra arise as a natural class of solutions to an associated set of difference equations (and an associated integral representation of the leading spectral density) that define the model. A special class of models, the Bessel models, is explored in section~\ref{sec:bessel-models}, containing important precursors of physics that will emerge in the full gravity theories to be developed later. Section~\ref{sec:multicritical} discusses properties  of the appropriate class of multicritical models that can be constructed as generalizations of the basic Wishart framework, as well as their double scaling limit which yields theories of gravity. These multicritical random matrix models are then used in section~\ref{sec:JT-gravity} to construct various models of JT supergravity, including the two crucial examples pertinent to BPS black holes: ${\cal N}{=}2$ and ${\cal N}{=}4$, the latter's multicritical description being presented here for the first time. Appendix~\ref{sec:gravity-embedding} reviews and discusses how the extended supergravity models arise in the black hole constructions (dressing the matrix model of the  pure JT supergravity construction with the required scale arising from the embedding).  Ample discussions of the results and ideas are presented throughout the paper as they are developed, and the paper ends with  some concluding remarks in section~\ref{sec:discussion}.

\section{Models of Positive Random Matrices}
\label{sec:models-of-RMM}

\subsection{Measure for Measure}
\label{sec:measure4measure}
We are interested in random matrix integrals where there is some number $\Gamma$ of degenerate zero energy states. Such systems are naturally modeled  using matrices of the form $M {=} Q^\dag Q$, where~$Q$ is a rectangular  matrix with $N$ rows and $(\Gamma + N)$ columns, with either real or complex  elements, drawn from an ensemble.  A matrix $Q$ is drawn from the ensemble with probability: 
\begin{equation}
\label{eq:probability}
  p(Q)=\frac{1}{{\cal Z}} {\rm e}^{-N{\rm Tr} V(M)}\ , \qquad  {\cal Z}\equiv\int {\rm e}^{-N{\rm Tr} V(M)} dQ \ ,
\end{equation} 
where $V(M)$ will be chosen to be polynomial in $M$, and ${\cal Z}$ is the partition function defined by integrating over all matrix elements of $Q$. The action has an invariance  under $Q\to U_L Q U_R$ where $U_{L,R}$ are independent unitary matrices, for~$Q$  complex. (Orthogonal matrices should be used instead if $Q$ is real.) Staying with complex $Q$ for now, the unitary transformation can be used to put $Q$  into a form where there is  an $N\times N$ block $\Lambda={\rm diag}\{y_1,y_2,\hdots,y_N\}$,
where $y_i$ are the non-zero eigenvalues, followed by $\Gamma$ columns of zeros, which   will yield the zero eigenvalues in the matrix~$M$. The integration measure for going from integrating over $Q$ to integrating over $\Lambda$ can be obtained   by computing the metric $\Tr |\delta Q|^2$ in terms of infinitesimal deformations along the independent non-trivial generators of $U_{L,R}$ (as reviewed in {\it e.g.} ref.~\cite{ForresterBook}). There are two key Jacobian factors: the expected Vandermonde determinant coming from the distinct eigenvalues above zero, as well as a factor coming from the degenerate sector. The partition function is\footnote{This form is sometimes referred to as the case $(\boldsymbol{\alpha},\boldsymbol{\beta}){=}(2\Gamma{+}1,2)$ in the Altland-Zirnbauer classification~\cite{Altland:1997zz}.} (after dropping a factor of the volume of the unitary group):
\be
\label{eq:dyson-gas-1}
{\cal Z}=\int_{-\infty}^\infty \prod_{i=1}^N d y_i \prod_{i < j} (y^2_i - y^2_j)^2 \prod_{i=1}^N y_i^{2\Gamma + 1} {\rm e}^{-N \sum_{i=1}^N V(y^2_i)} \ .
\ee
It is also instructive to write the partition function in terms of the $N$ non-zero {\it positive} eigenvalues $\lambda_i=y_i^2$ and the $\Gamma$ zero eigenvalues of the matrix $M$, yielding, for real or complex $Q$: 
\be
\label{eq:dyson-gas-2}
{\cal Z}=\int_{0}^\infty \prod_{i=1}^N d \lambda_i \prod_{i < j} |\l_i - \l_j|^\beta \prod_{i=1}^N \lambda_i^{\frac{\beta}{2} (\Gamma + 1)-1} {\rm e}^{-N \sum_{i=1}^N V(\lambda_i)} \ .
\ee
In the above $\beta=1,2$ corresponds to real/complex entries for $Q$.  From the point of view~(\ref{eq:probability}) of drawing from a random ensemble, this diagonalization now results in the probability of draw of a list of the  eigenvalues $\{{\bf 0},\lambda_1,\lambda_2,\ldots,\lambda_N\}$ (here ${\bf 0}$ denotes the list of $\Gamma$ zeros):
\begin{equation}
\label{eq:probability2}
p(\lambda_1,\lambda_2,\ldots,\lambda_N)=
    \frac{1}{\cal Z}  \prod_{i < j} |\l_i - \l_j|^\beta \prod_{i=1}^N \lambda_i^{\frac{\beta}{2} (\Gamma + 1)-1} {\rm e}^{-N \sum_{i=1}^N V(\lambda_i)} \ .
\end{equation}
We will consider the complex case ($\beta{=}2$) henceforth, but our results will apply more  generally. Models of this kind with general $V(M)$ were first studied in the context of string theory in refs.~\cite{Morris:1991cq,Morris:1990bw,Dalley:1992qg,Dalley:1992vr}, and more recently used for studies of various dilaton (JT) gravity theories in {\it e.g.,} refs.~\cite{Johnson:2019eik,Johnson:2020heh,Johnson:2020exp,Johnson:2021owr,Johnson:2024fkm}.

\subsection{A Choice: Degenerate or Non-Degenerate $M$} 
\label{sec:degenerate_or_not}
It is important to note that in constructing the matrix $M$ we actually had two distinct choices, either  $M{=}Q^\dag Q$, or $M{=}Q Q^\dag$. In our conventions above, the first choice amounts to an $(N{+}\Gamma){\times}(N{+}\Gamma)$ matrix with $\Gamma$  eigenvalues that {\it must} be zero. These will play the role of BPS states in what is to come. The remaining  $N$ eigenvalues have no such constraint. On the other hand, the second choice gives an $N\times N$ matrix with $N$ generic  eigenvalues, and no special degenerate states (and hence no BPS states). In the  literature, the difference corresponds to whether the ratio $(N+\Gamma)/N$ is greater or less than unity. The measure above is the same for either choice, and to convert between normalizations  for each case the potential can be rescaled to have either $N$  or $(N+\Gamma)$ in front.\footnote{Sometimes, the sign of $\Gamma$ is used as a shorthand for the difference between cases. Positive~$\Gamma$ is the  case with~$\Gamma$ zeros, while negative $\Gamma$ is the other case. This is studied in detail in ref.~\cite{Johnson:2021rsh} for Bessel models both analytically and explicitly by constructing the two cases as random matrix ensembles numerically. Note that the sign conventions for $\Gamma$ are reversed there compared to here. We will mostly try to avoid thinking in terms of the sign of $\Gamma$ for this paper since, if used in the measure, it can lead to some confusion. We thank Maciej Kolanowski for discussions about this.}
It is  useful to consider both possibilities, although most of this paper will have the original ($\Gamma$ degenerate zeros) situation in mind. 
In either case,  in the Dyson gas picture to be discussed in section~\ref{sec:dyson-gas}, the measure  typically generates an effective extra logarithmic repulsion away from $\lambda{=}0$ for the continuum of~$N$ eigenvalues of $Q$. Note that the eigenvalue repulsion mechanism is best thought of in terms of $Q$, the underlying random matrix, rather than the matrix $M$ formed from it, although sometimes the language is used for both.

In both cases, 
the edge of the continuum part of the leading spectrum will start at non-zero energy. In the case with $\Gamma$ degenerate zero energy states for $M$ there will also be a $\delta$-function contribution at the origin proportional to $\Gamma/N$, in our conventions. This situation will be  interpreted as having  a gap in the spectrum. When there are no zero energy states, this situation is instead interpreted as the spectrum simply starting at some shifted value of the energy.  Both types of behaviour will appear naturally later when constructing  random matrix descriptions for gravitational systems. 

\subsection{The Wishart Prototype and Generalizations}
\label{sec:wishart-and-more}
%
%
\begin{figure}
    \centering
    \includegraphics[width=0.45\textwidth]{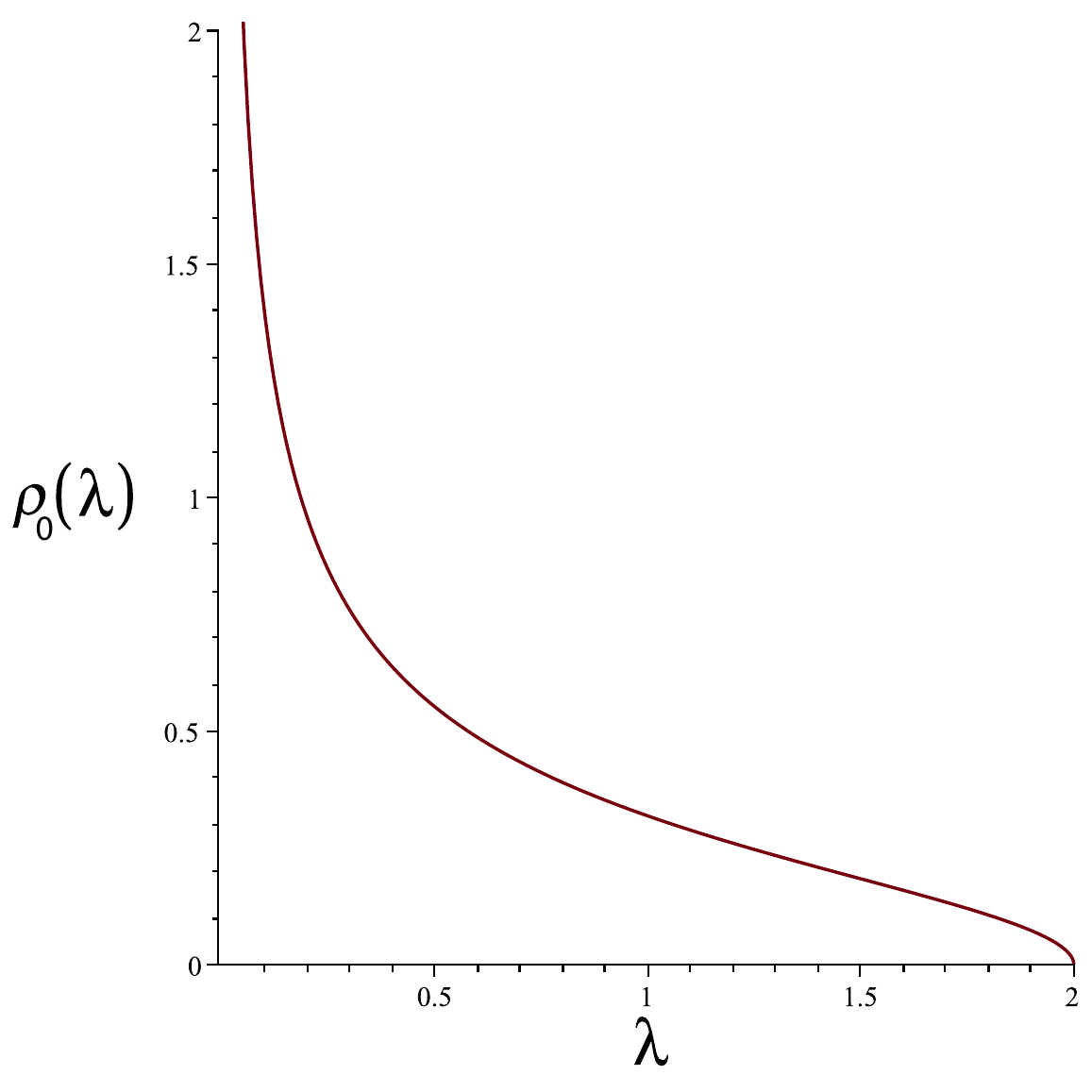}
  \caption{\rm The leading density~(\ref{eq:density-basic})  for the simplest Wishart model. There is a divergence at the origin going as $\rho_0(\l) \sim \lambda^{-1/2}$.}
    \label{fig:0-1-density-and-potential}
\end{figure}
The simplest case of linear potential gives   the classic Wishart model,  sometimes known as the Wishart-Laguerre ensemble for reasons to unfold later. 
For orientation, in the case $\Gamma{=}0$, the leading  large~$N$ dynamics
 of the Dyson gas~(\ref{eq:dyson-gas-1}) is identical~\cite{Dalley:1992qg} to that of a symmetric Hermitian matrix model,  and with potential $V(y){=}2y^2$, the leading spectral density is the classic Wigner semi-circle with endpoints at $y=\pm\sqrt{2}$:
\begin{equation}
    \rho_0(y)=\frac{1}{2\pi}\sqrt{2-y^2}\ , 
\end{equation}
which on transforming to the system~(\ref{eq:dyson-gas-2}) on the half--line with $V(\lambda)=2\lambda$, transforms into a spectral density $\rho_0(\lambda)$ with a square root divergence at the origin:
\begin{equation}
\label{eq:density-basic}
    \rho_0(\lambda)=\frac{1}{\pi}\frac{\sqrt{2-\lambda}}{\sqrt{\lambda}}\ .
\end{equation}
It is plotted in figure~\ref{fig:0-1-density-and-potential}. It is the behaviour at the left endpoint (the origin) that will be of most interest in what follows, and two natural departures from this simple case immediately suggest themselves:

\begin{enumerate}[{\bf (A)}]
    \item One is to choose the potential $V(\lambda)$ in order to allow the endpoint to have more general ``multicritical" behaviour, where the spectral density develops $k$ zeros at the endpoint, allowing for a description of smooth macroscopic surfaces and hence 2D gravity. This was pioneered and explored long ago, starting with  refs.~\cite{Morris:1991cq,Morris:1990bw,Dalley:1992qg} (here $\Gamma$  denotes the $\Gamma$-function):
\begin{equation}
\label{eq:multicritical}
\rho_0^{(k)}=(-1)^k\frac{(k+1)!\Gamma\!\left(\frac12-k\right)}{2^k\pi^{3/2}}\lambda^{k-\frac12} \sqrt{2-\lambda}\ .
\end{equation}
\item The other is to turn on $\Gamma$.  If $\Gamma$ scales with $N$, this will result in a gap in the spectrum at leading order. This gives the Marchenko-Pastur~\cite{Pastur:1967zca} distribution, which is (writing $\Gamma/N={\widetilde\Gamma}$):
\begin{equation}
\label{eq:marchenko-pastur}
    \rho^{\rm MP}_0(\lambda) =  
   \frac{1}{\pi\lambda} {\sqrt{(\lambda_+-\lambda)(\lambda-\lambda_-)}}\; \qquad \lambda_\pm=\frac{\widetilde\Gamma}{2}+1\pm\sqrt{{\widetilde\Gamma}+1}
\end{equation}
which will be derived in two ways in the next subsection. Intuitively, in the Dyson gas picture, in addition to the eigenvalue repulsion force arising from the Vandermonde determinant involving the~$N$ eigenvalues $\lambda_i$, there is also a repulsion coming from the term $\lambda^{\Gamma}$, corresponding to the effects of the $\Gamma$ eigenvalues at the origin. For generic $\Gamma$, in the large $N$ limit this repulsion is subdominant, and the leading spectral density is unchanged, but if $\Gamma$ scales with $N$, those eigenvalues conspire to push the endpoint away from zero to $\lambda=\lambda_-$. 
\end{enumerate}

The multicritical points of {\bf (A)}, themselves extended to incorporate $\Gamma$ (as was first  done in ref.~\cite{Dalley:1992br}) give rise to, after a double scaling limit, non-trivial models of two-dimensional gravity that are useful as building blocks for random matrix descriptions of supersymmetric gravity such as JT supergravity. Working with {\it both} non-zero $k$ and non-zero $\Gamma$ allows for the study of several important JT supergravity  models, as shown in refs.~\cite{Johnson:2020heh,Johnson:2020exp,Johnson:2021owr} for ${\cal N}=1$ and ref.~\cite{Johnson:2023ofr} for ${\cal N}=2$. Later in this paper it will be shown that the case of ${\cal N}=4$ can be studied this way as well. Note that in all these models, the limit where~$\Gamma$ scales with $N$ must be taken {\it after} the double scaling limit around the $\lambda=0$ endpoint. This will be discussed below.

\subsection{Large $N$ Analysis, I - Dyson Gas}
\label{sec:dyson-gas}
We begin by giving a general argument that introducing a number $\G$ of ground states that scales with~$N$ will introduce a gap in the spectral density of order $\G/N$ that depends on the fine details of the matrix potential. Recalling the partition function \eqref{eq:dyson-gas-2} 
\be
\label{eq:partition-function}
{\cal Z} = \int_0^\infty \prod_{i=1}^N d\lambda_i \prod_{i< j}(\l_i -\l_j)^2 \prod_i  \l_i^\Gamma {\rm e}^{-N \sum_i^N V(\l_i)}\ ,
\ee
we would like to find the large $N$ distribution for the eigenvalues $\l_i$. The saddle-point equation for an individual eigenvalue is given by
\be
N V'(\l_i) - \frac{\G}{\l_i} - 2 \sum_{j\neq i} \frac{1}{\l_i-\l_j}=0.
\ee
In the large $N$ limit the full density of states will consist of $\Gamma$ eigenvalues at zero, along with a continuum of eigenvalues within a range $\l \in (\l_-,\l_+)$. We are interested in the form of the density of states away from $\l=0$. Let us thus define the density of states excluding the $\Gamma$ ground states  by $\rho(\l)=\frac{1}{N} \sum_{i=1}^N \delta(\l-\l_i)$.  It is normalized according to: $\int \rho(\l)d\lambda=1$. Treating the eigenvalue $\l$ as a general complex variable $z$, we can rewrite the saddle-point equation as
\be 
V'(z) = \frac{\Gamma}{ N} \frac{1}{z}+ 2 \dashint_{-\infty}^\infty d \l \frac{\rho(\l)}{z-\l} \,,
\ee
where the bar on the integral sign indicates the principal value integral. Setting $\Gamma{=}0$ results in the standard saddle-point equation for the density of states in matrix integral calculation. It is useful to introduce the resolvent, defined for complex $z$ through:
\be \label{eqn:resolvent}
G(z) = \frac{1}{N}\sum_{i=1}^N \frac{1}{z-\l_i} \equiv \dashint d \l \frac{\rho(\l)}{z - \l} \,,
\ee
where all integrals are defined along the real axis. In the complex plane the resolvent should be analytic away from the branch cut where the spectrum has support.  When $z \in (\l_-,\l_+)$ which are the endpoints of the cut, we have:\footnote{Here we have made use of the representation: $\lim_{\epsilon \to 0}\frac{1}{\pi}\frac{\epsilon}{x^2 + \epsilon^2} = \delta(x).$}
\be
G(z\pm i \epsilon) = \dashint d \l \frac{\rho(\l)}{z - \l} \mp i \pi \rho(z)\ .
\ee
Rewriting the saddle-point equation we have:
\begin{gather}
\rho (z) = \frac{1}{2\pi i} \lr{G(z-i \epsilon)-G(z+i \epsilon)}\,,\\
G(z\pm i \epsilon) = \frac{1}{2} \lr{V'(z) - \frac{\Gamma}{N} \frac{1}{z}} \mp i \pi \rho(z) \,.
\end{gather}
Combining the above equations, along with taking the $z{\to}\infty$ limit of \eqref{eqn:resolvent} we find that three properties must be satisfied:
\begin{gather} \label{eqn:resolvent_conditions}
    \lim_{z\to \infty} G(z) = \frac{1}{z} \,,\\
    \rho(\l) = \frac{1}{2\pi i} \lr{ G(\l-i\epsilon) - G(\l+i \epsilon) } \,, \\ 
    G(z-i\epsilon) + G(z+i \epsilon) = V'(z) - \frac{\Gamma}{N} \frac{1}{z} \ ,
\end{gather}
Similar to the case without ground states states, we can take an ansatz:
\be
G(z) = \frac{1}{2} \lr{V'(z)-\frac{\Gamma}{N}\frac{1}{z}} - P(z) \sqrt{(z-\l_-)(z-\l_+)}\,.
\ee
Let us take $V'=2$, matching our earlier convention. Demanding $z^{-1}$ fall-off at infinity we find $P(z) = \frac{1}{ z}$. The continuum density of states is thus given by:
\be
\label{eq:leading-density-dyson-approach}
\rho(\lambda) =  \frac{1}{\pi} \frac{1}{ \lambda} \sqrt{(\l-\lambda_- )(\lambda_+ - \lambda)}\ ,
\ee
and there is a $\frac{\Gamma}{N}\delta(0)$ at the origin when $(\Gamma+N){>}N$.
We have two additional conditions to impose: the integrated density $\rho(\lambda)$ must be unity and the resolvent must fall off as $G = z^{-1} + \mathcal{O}(z^{-2})$ with unit coefficient as $z\to \infty$. This gives us two equations for the two unknown endpoints $\l_{\pm}$:
\begin{gather}
\frac{1}{2} \lr{\l_- + \l_+ -2 \sqrt{\l_- \l_+}}=1\ , \qquad \ 
\frac{1}{2}(\l_+ + \l_-) - \frac{\Gamma}{2 N} = 1\ .
\end{gather}
The solution for the cuts, setting the endpoint of the eigenvalue distribution, is given by:
\be
\lambda_\pm = \frac{1}{2} \lr{1 \pm \sqrt{\frac{\Gamma}{N}+1}}^2\ .
\ee
These expressions will be derived in a different way in the following subsection. We see that scaling $\Gamma \sim \mathcal{O}(N)$ will give a gap in the large $N$ limit away from $\l=0$ as expected.



\subsection{Large $N$ Analysis, II - Orthogonal Polynomials}

\label{sec:orthog-poly}
A more powerful approach to the matrix model is to work in terms of a family of orthogonal polynomials, $P_n(\lambda)=\lambda^n+\mbox{\rm lower powers}$, defined on the half line such that
\begin{equation}
\label{eq:orthogonality}
    \int_0^\infty  P_n(\lambda)P_m(\lambda) \lambda^\Gamma {\rm e}^{-NV(\lambda)} d\lambda = h_n\delta_{mn}\ ,
\end{equation}
where $h_n$ are numbers arising from having normalized the highest power of $\lambda$ to unity in a polynomial, and we define $h_0{\equiv}\int_0^\infty d\mu(\lambda)$. Orthogonal polynomials with respect to a quite general measure $d\mu(\lambda)=w(\lambda)d\lambda$ (with  well-behaved enough\cite{szego1975orthogonal,chihara} weight function $w(\lambda)$) satisfy a three term recursion relation:
\begin{equation}
\label{eq:recursion}
    \lambda P_n(\lambda)=P_{n+1}(\lambda)+S_nP_n(\lambda)+R_nP_{n-1}(\lambda)\ ,
\end{equation}
where the fact it truncates after three terms follows from orthogonality.  Note that multiplying~(\ref{eq:recursion}) by $R_{n+1}$ and using~(\ref{eq:orthogonality}) yields the relation $R_n{=}h_n/h_{n-1}$, which will be useful below.  

Consider forming the matrix $\|P_i(\lambda_j)\|$, whose $i$th row and $j$th column entry is the polynomial $P_i(\lambda_j)$.  The vandermonde determinant $\prod_{i<j}(\lambda_i-\lambda_j)$ can be written as ${\rm det}\|P_i(\lambda_j)\|$ by taking suitable linear combinations of  columns. It follows from this that  matrix model quantities can be expressed entirely in terms of the $P_n$, in particular the first $N$ of them. For example, the probability $p(\lambda)d\lambda$ of finding a single {\it non-zero} eigenvalue in some interval $(\lambda,\lambda+d\lambda)$ can be derived by integrating the basic probability~(\ref{eq:probability}) over the $N-1$ other eigenvalues $\lambda_i$, whereupon $p(\lambda)$
 can be written as:
 \begin{equation}
 \label{eq:density-ancestor}
     p(\lambda)= \lambda^\Gamma{\rm e}^{-NV(\lambda)} \frac{1}{N}\sum_{n=0}^{N-1}\frac{1}{h_n}P_n^2(\lambda)\ .
 \end{equation}
(See {\it e.g.,} Appendix 4 of ref.~\cite{Bessis:1980ss} for example derivations.)

The key point here is that the matter of solving the matrix model boils down to finding the recursion coefficients $S_n$ and $R_n$, which ultimately define the $P_n(\lambda)$. As a familiar example of a set of polynomials, the simple  Wishart case mentioned above (with linear potential) defines (up to a convention) the  case of the (generalized) Laguerre polynomials, familiar from the study of quantum aspects of the Hydrogen atom. (This is entirely analogous to the case of Gaussian Hermitian matrix models, with eigenvalues on the whole real line, where the orthogonal polynomials that emerge are the Hermite polynomials.)

\subsubsection{The Defining Equations}
Equations for the recursion coefficients  can be straightforwardly derived (as was done in {\it e.g.}, refs.~\cite{Morris:1991cq,Dalley:1992qg}) by considering the following three identities: 
\begin{eqnarray}
    \int_0^\infty d\mu(\lambda) \,\,P_n(\lambda)\frac{d}{d\lambda} P_n(\lambda)  &=& 0\ ,\label{eq:identity1}\\ 
    \int_0^\infty d\mu(\lambda) 
    \,\,P_{n-1}(\lambda)\frac{d}{d\lambda} P_n(\lambda)  &=& nh_{n-1}\ , \label{eq:identity2}\\
    \int_0^\infty d\mu(\lambda)
    \,\,P_n(\lambda) \lambda \frac{d}{d\lambda} P_n(\lambda)  &=& nh_n\label{eq:identity3}\ ,
\end{eqnarray}
where $d\mu(\lambda){\equiv}\lambda^\Gamma {\rm e}^{-NV(\lambda)} d\lambda$. The next step is to integrate by parts. How things work in detail depends upon whether $\Gamma$ vanishes or not, although the final expression can be put into a form that uniformly applies to both cases. 

Take first the case of $\Gamma{=}0$. When the derivative hits the exponential part of the measure, it will produce a term with an insertion of $-NV^\prime(\lambda)$ into the integral containing the two polynomials. There will also be non-zero boundary terms coming from evaluating the polynomials at $\lambda{=}0$ (the other boundary contribution vanishes because of the exponential). Introducing an extra bit of notation, the results can be written as follows:
\begin{eqnarray}
    \langle n|V^\prime({\lambda})|n\rangle &=& \frac{P^2_n(0)}{Nh_n}\ ,\label{eq:newidentity1}\\ 
    \langle n-1|V^\prime({\lambda})|n\rangle   &=& \frac{n}{N\sqrt{R_n}}+\frac{P_{n-1}(0)P_{n}(0)}{N\sqrt{h_nh_{n-1}}} \ , \label{eq:newidentity2}\\
    \langle n|{\lambda} V^\prime({\lambda})|n\rangle  &=& \frac{1}{N}(2n+1)\label{eq:newidentity3}\ ,
\end{eqnarray}
where the compact notation 
$ |n\rangle\equiv \frac{P_n(\lambda)}{\sqrt{h_n}}$ means~\cite{Bessis:1980ss,Gross:1990aw}  the orthogonality integral~(\ref{eq:orthogonality}) rescaled to give a unit normalized inner product $\langle n|m\rangle=\delta_{mn}$. The recursion relation \eqref{eq:recursion} in terms of this notation is given by $\l |n \rb = \sqrt{R_{n+1}} | n+1 \rb + S_n | n \rb + \sqrt{R_n} | n-1 \rb$.

The above equations can be rewritten in a useful form where they only contain inner products of $V'$. Combining the first two identities \eqref{eq:newidentity1}-\eqref{eq:newidentity2} by eliminating the boundary terms on the right-hand side we get
\be
\lb n | V' | n \rb \lb n-1 | V'| n-1 \rb = \lr{\lb n | V' | n\rb - \frac{n}{N \sqrt{R_n}}}^2\,.
\ee
The third identity \eqref{eq:newidentity3} can be rewritten involving only inner products of $V'$ using the recursion \eqref{eq:recursion}
\be
\sqrt{R_{n+1}} \lb n | V' | n+1\rb + S_n \lb n | V' | n\rb + \sqrt{R_n} \lb n | V' | n-1\rb = \frac{1}{N}(2n+1)\,.
\ee
As we will see momentarily these two equations contain all information about the matrix integral. The two analogous simple expressions with $\Gamma$ turned on will be written below (the reader can skip ahead to equations~(\ref{eq:the-omegas}) and~(\ref{eq:full-string-equations-discrete}) if desired). 

Consider now the case of $\Gamma\neq0$.  The procedure of integrating by parts is somewhat different. In this case the total derivative vanishes giving us the conditions
\begin{gather}
    \int_0^\infty \frac{\partial}{\partial \l} \lr{P_n(\lambda) P_n(\lambda) \lambda^\Gamma {\rm e}^{-NV(\lambda)}}  d\lambda = 0\,,\\
    \int_0^\infty \frac{\partial}{\partial \l} \lr{P_{n-1}(\lambda) P_n(\lambda) \lambda^\Gamma {\rm e}^{-NV(\lambda)}}  d\lambda = 0\,,\\
    \int_0^\infty \frac{\partial}{\partial \l} \lr{\l P_n(\lambda)  P_n(\lambda) \lambda^\Gamma {\rm e}^{-NV(\lambda)}}  d\lambda = 0\,,
\end{gather}
since the boundary terms are killed by the factor of $\l^\Gamma$. Let us explicitly work out the first identity which reduces to
\be
-N \int_0^\infty d \mu  V'(\l) P_n(\lambda) P_n(\lambda) + \Gamma \int_0^\infty d \mu \frac{P_n(\l) P_n(\l)}{\l}=0\,.
\ee
The second term is slightly unusual, and can be handled in a number of ways~\cite{Anderson:1991ku,chihara,Myers:1991akt,Klebanov:2003wg} depending on the desired final form. Following instead the alternative approach given in {\it e.g.,} ref.~\cite{Klebanov:2003wg} we notice we can expand\footnote{Expanding in a series in $\l$, the most singular term must be $P_n(0)/\l$, with all higher order terms expandable in orthogonal polynomials. Since $P_1 = 1$ we have that the constant term is $P_n'(0)$, with all remaining terms proportional to higher order polynomials that vanish when used in the integral.}
\be
\frac{P_n(\l)}{\l} = P_{n-1}(\l) + \ldots + P_n'(0) + \frac{P_n(0)}{\l}\,.
\ee
Note that when we plug this expansion back into the second term in the integral, by orthogonality all contributions except the last one vanish. Iterating this procedure for both copies of $P_n$ in the second term, we would pick up the second to last term as well since it is proportional to $P_n'(0) \propto P_0 = 1$. The identity simplifies to 
\be
-N \int_0^\infty d \mu  V'(\l) P_n(\lambda) P_n(\lambda) + \Gamma \lr{P^2_n(0) h_0^{[\Gamma-1]}+P_n(0)P_n'(0) h_0} = 0\,.
\ee 
In the second term we have used that $\l^{-1} d \mu$ is the measure for the closely related  family of orthogonal polynomials defined for $\Gamma{-}1$, with $h_0^{[\Gamma-1]} = \int \l^{-1}d \mu$. The other conditions follow from similar manipulations, and we arrive at the identities in our previous notation
\begin{eqnarray}
    \langle n|V^\prime({ \lambda})|n\rangle &=& \frac{\Gamma}{Nh_n}\left(P^2_n(0)h_0^{[\Gamma-1]}+P_n(0)P^\prime_n(0)h_0\right)\ ,\label{eq:newidentity1b}\\ 
    \langle n-1|V^\prime({ \lambda})|n\rangle   &=& \frac{n}{N\sqrt{R_n}}+\frac{\Gamma}{N\sqrt{h_nh_{n-1}}}\left(P_n(0)P_{n-1}(0)h_0^{[\Gamma-1]}+P_{n-1}(0)P^\prime_{n}(0)h_0 \right)  \label{eq:newidentity2b}\\
    &=& \frac{n}{N\sqrt{R_n}}+\frac{\Gamma}{N\sqrt{h_nh_{n-1}}}\left(P_{n}(0) P_{n-1}(0)h_0^{[\Gamma-1]}+P_{n}(0)P^\prime_{n-1}(0)h_0+h_{n-1}\right) \ , \nonumber\\ 
    \langle n|{\lambda} V^\prime({ \lambda})|n\rangle  &=& \frac{1}{N}(2n+1+\Gamma)\label{eq:newidentity3b}\,.
\end{eqnarray}
Note that there are two different presentations of the second identity, coming from expanding either $\l^{-1} P_n(\lambda)$ or $\l^{-1} P_{n-1}(\lambda)$ as done for the first identity. 

The $P_n(0)$, $P^\prime_n(0)$, {\it etc.,} can now be eliminated in much the same way as done for $\Gamma=0$, giving a pair of relations that generalize the $\Gamma{=}0$ case. We define the quantities~\cite{Dalley:1991qg}:
\begin{equation}
\label{eq:the-omegas}
    \Omega_n \equiv \langle n|V^\prime({\lambda})|n\rangle\ , \quad {\rm and}\quad {\widetilde\Omega}_n  \equiv \frac{n}{N}-\sqrt{R_n}\langle n-1|V^\prime({\lambda})|n\rangle\ ,
\end{equation}
in terms of which we can write the final result  in a unified way for both the $\Gamma{=}0$  and $\Gamma{\neq}0$ cases: 
\begin{eqnarray}
\label{eq:full-string-equations-discrete}
    R_n\Omega_n\Omega_{n-1} &=& {\widetilde\Omega}_{n}^2+\frac{\Gamma}{N}{\widetilde\Omega}_{n}\nonumber\\
    {\widetilde\Omega}_{n}+{\widetilde\Omega}_{n+1}&=&S_n{\Omega}_{n}-\frac{\Gamma}{N}\ .
\end{eqnarray}
In the first line, it is helpful to use one of each version of identity~(\ref{eq:newidentity2b}) in forming ${\widetilde\Omega}_n^2$, as well as the relation $R_n=h_n/h_{n-1}$. The second line follows from using the recursion~(\ref{eq:lambda-operator}) acting to the left of equation~\eqref{eq:newidentity3b}.

In the large $N$ limit, the combination $X=\frac{n}{N}$ defines a continuous variable on the interval $[0,1]$, and defining also $\epsilon=\frac{1}{N}$, quantities with indices $n$ become functions of $X$, with shifts by $\pm1$ translating to  shifts by $\pm\epsilon$, hence. The results~(\ref{eq:full-string-equations-discrete}) can then be written as:
\begin{eqnarray}
\label{eq:full-string-equations-large-N}
    R(X)\Omega(X)\Omega(X-\epsilon) &=& {\widetilde\Omega}(X)^2+\epsilon\Gamma{\widetilde\Omega}(X)\nonumber\\
    {\widetilde\Omega}(X)+{\widetilde\Omega}(X+\epsilon)&=&S(X){\Omega}(X)-\epsilon\Gamma\ .
\end{eqnarray}
It is useful to look at the leading large $N$ form of this result, coming from dropping $\epsilon=1/N$.
The terms involving $\Gamma$  generically disappear at leading order, unless $\Gamma$ scales with~$N$. So writing $\Gamma={\widetilde\Gamma}N$, taking the leading large $N$ terms gives the equations:
\begin{eqnarray} \label{eqn:largeN_String_eqn}
    R(X) = \frac{{\widetilde\Omega}(X)^2}{\Omega(X)^2} + {\widetilde\Gamma}\frac{{\widetilde\Omega}(X)}{\Omega(X)^2}\ ; \qquad 
S(X) = 2\frac{{\widetilde\Omega}(X)}{\Omega(X)} + {\widetilde\Gamma}\frac{1}{\Omega(X)}\ .
\end{eqnarray}

\subsubsection{The Leading Spectral Density as an Integral} 

A very useful relationship between the large $N$ orthogonal polynomial quantities $R(X)$ and $S(X)$ and the leading spectral density was derived in ref.~\cite{Dalley:1990hb} and explored in refs.~\cite{Dalley:1991zs,Dalley:1991jp} (it extends the result of~\cite{Bessis:1980ss} for purely even potentials).  That work had $\Gamma=0$ but here it is natural to explore the effects of large non-zero $\Gamma={\tilde \Gamma}N$. The eigenvalue density $\rho(\lambda)$ is the large $N$ limit of the single eigenvalue probability~(\ref{eq:density-ancestor}). Hence, the moments  of $\rho(\lambda)$ can be naturally expressed in the orthogonal polynomial (and inner product) language  as:
\begin{equation}
    \label{eq:moments1}
    \int d\lambda \rho(\lambda)\lambda^p = \lim\limits_{N\to\infty}
    \frac{1}{N}\sum_{n=0}^{N-1} \int d\mu(\lambda)\frac{1}{h_n} P^2_n(\lambda) \lambda^p=
    \lim\limits_{N\to\infty}
    \frac{1}{N}\sum_{n=0}^{N-1}\langle n |{\hat\lambda}^p|n\rangle\,.
\end{equation}
Defining raising and lowering operators such that ${\cal A}^\dagger|n\rangle=|n+1\rangle$ and ${\cal A}|n\rangle=|n-1\rangle$, multiplication by $\lambda$ inside the integral, combined with using recursion relation~(\ref{eq:recursion}), becomes the insertion of the operator:\footnote{Note that we also have $\lb n | \mathcal{A} = \lb n+1|, ~ \lb n | \mathcal{A}^\dag = \lb n-1|$.}
\begin{equation}
    \label{eq:lambda-operator}
    {\hat\lambda}\equiv \sqrt{R_{n+1}}{\cal A}^\dagger+\sqrt{R_n} \mathcal{A} + S_n\ ,
\end{equation}
with the commutation relations
\begin{equation}
\label{eq:operator-rules}
    {\cal A}^\dagger\sqrt{R_n}=\sqrt{R_{n+1}}\,{\cal A}^\dagger\ ,   \mbox{\rm and}\quad  {\cal A}\,\sqrt{R_n}=\sqrt{R_{n-1}}\,{\cal A} \,.
\end{equation}

Extracting the leading large $N$ piece of this is done straightforwardly by simply ignoring the ordering of terms in $\hat\lambda$ and keeping track of the number of  terms that have the same number of ${\cal A}^\dagger$ and $\cal A$ occurrences - this is all counted by binomial coefficients:
\begin{eqnarray}
    \label{eq:moments2}
     \int d\lambda \rho_0(\lambda)\lambda^p =
    \lim\limits_{N\to\infty}
     \frac{1}{N}\sum_{n=0}^{N-1}\langle n |
     \sum_{r=0}^{p}\frac{p!}{(p-r)!r!}S^{p-r}_n
     (\sqrt{R_{n+1}}{\cal A}^\dagger+\sqrt{R_n} {\cal A} )^r|n\rangle\nonumber\\
     \approx\lim\limits_{N\to\infty}
     \frac{1}{N}\sum_{n=0}^{N-1}
     \sum_{r=0}^{p}\frac{p!}{(p-r)!r!} \times \frac{r!}{(r/2)!(r/2)!} S^{p-r}_n
     R_n^{r/2}
     \ ,
\end{eqnarray}
where in the last line $r$ really must be even in the sum, and we have approximated that $R_{n+1}\sim R_n$ at large $N$. The following identity:
\begin{equation}
    \frac{r!}{(r/2)!(r/2)!} = \int^{1}_{-1}\frac{dy}{\pi}\frac{(2y)^r}{\sqrt{1-y^2}}\ ,
\end{equation}
 allows a  re-writing, and after re-summing (over both even and odd $r$ cases; the latter give zero from the identity):
\begin{equation}
    \int d\lambda \rho_0(\lambda)\lambda^p 
    =\lim\limits_{N\to\infty}
     \frac{1}{N}\sum_{n=0}^{N-1}
     \int_{-1}^{1}\frac{1}{\pi}\frac{dy}{\sqrt{1-y^2}}(2y\sqrt{R_n}+S_n)^p 
     \ ,
\end{equation}
from which it follows that:
\begin{equation}
    \rho_0(\lambda) = \frac{1}{\pi}\int_0^1 dX
    \int_{-1}^{1} \frac{dy}{\sqrt{1-y^2}}\delta\bigl(\lambda - [2y\sqrt{R(X)}+S(X)]\bigr)\ ,
\end{equation}
where we have taken the large $N$ limit and used the continuous coordinate $X=\frac{n}{N}$ as before. Doing the $y$-integral leaves a useful integral representation of the leading density: 
\begin{equation}
\label{eq:integral-representation}
    \rho_0(\lambda) = \frac{1}{\pi}\int_0^1
dX\frac{\Theta\left[4R(X)-(\lambda-S(X))^2\right]}{\sqrt{4R(X)-(\lambda-S(X))^2}}\,.
\end{equation}
For a given $\lambda$, there are two solutions for vanishing of the square root in the denominator, corresponding to the rising and falling branches of $\rho_0(\lambda)$. The density falls to its  smallest value when $S(X)$ and $R(X)$ rise to their largest, which is at $X=1$.  Therefore $\rho_0(\lambda)$ has support in the range $\l \in (\l_-,\l_+)$ with
\be \label{eqn:eigenvalue_endpoints}
\lambda_\pm=\left.\lr{S(X)\pm2\sqrt{R(X)} }\right\rvert_{X=1}\,.
\ee
The center of the distribution is at
$S_c\equiv S(X=1)$, with width $4\sqrt{R_c}\equiv 4\sqrt{R(X=1)}$.

\begin{figure}
    \centering
    \includegraphics[width=0.45\textwidth]{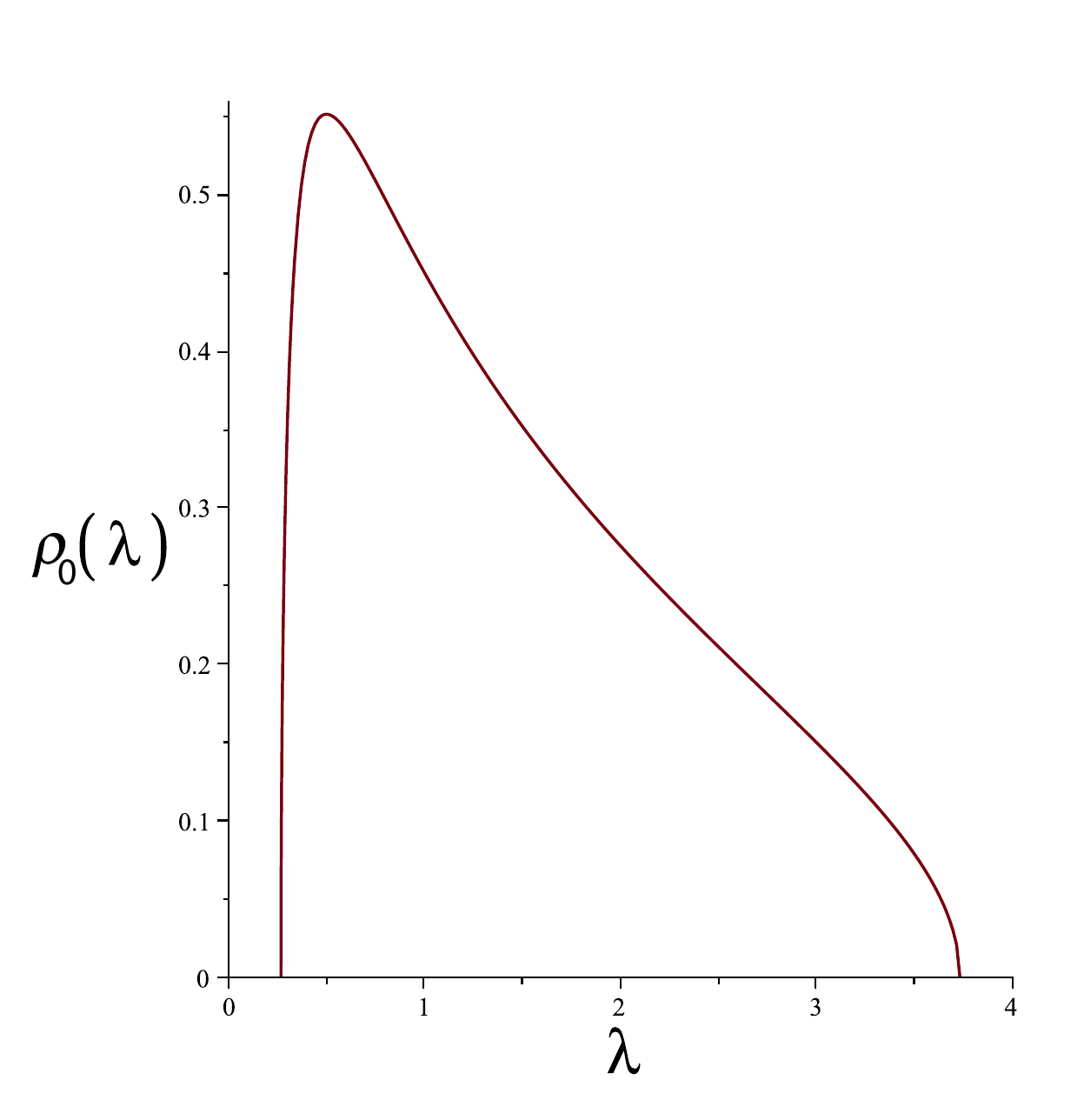}
    \caption{The Marchenko-Pastur density~(\ref{eq:unscaled-gaps}) for the case ${\widetilde\Gamma}{=}2$, showing the gap: $\lambda_-{=}2{-}\sqrt{3}$. Although not indicated, there are two eigenvalue $\delta$-functions localized at $\l{=}0$  not explicitly included in  formula~(\ref{eq:unscaled-gaps}).}
    \label{fig:MP-density}
\end{figure}

This is a good point to revisit the classic  Wishart example, where $V(\lambda)=2\lambda$. In this case, things simplify dramatically since $V^\prime=2$, and so from~(\ref{eq:the-omegas}), the $\Omega_n=2$ and  ${\widetilde\Omega}_n=n/N$, yielding, from~(\ref{eq:full-string-equations-discrete}):
\begin{equation}
\label{eq:laguerre-coeffs}
    S_n=\frac{2n+1+\Gamma}{2N}\ , \quad\mbox{\rm and}\quad R_n=\frac{n(n+\Gamma)}{4N^2}\ ,
\end{equation}
which can be recognized as defining the recursion for the (generalized) Laguerre polynomials~\footnote{The Laguerre polynomials $L_n^{\alpha}(\lambda)$ are often normalized such that $L_n^{\alpha}(\lambda)=(-)^n\lambda^n/n!+\cdots$, so the connection is made after normalizing them so that they start as $\lambda^n$. The factors $2N$ and $(2N)^2$ in the denominators in (\ref{eq:laguerre-coeffs}) would be replaced by 1 in the comparison, since for Laguerre the measure is $ \lambda^\alpha{\rm e}^{-\lambda}d\lambda$. The parameter $\alpha$ is our parameter $\Gamma$ here.}. Taking the large $N$ limit and holding ${\widetilde\Gamma}=\Gamma/N$ fixed yields:
\begin{equation}
S(X)=X+\frac{\widetilde\Gamma}{2}\ , \quad\mbox{\rm and}\quad R(X) =\frac{X^2}{4}+\frac{\widetilde\Gamma}{4}X\ ,
\end{equation}
which upon substitution into~(\ref{eq:integral-representation}), yields:
\begin{eqnarray}
\label{eq:unscaled-gaps}
    \rho_0(\lambda) = \frac{1}{\pi}\int^1_0 dX
    \frac{\Theta\left[2\lambda X -\left(\frac{\widetilde\Gamma}{2}-\lambda\right)^2\right]}{\sqrt{2\lambda X - \left(\frac{\widetilde\Gamma}{2}-\lambda\right)^2}}=\frac{1}{\pi}
    \frac{\sqrt{(2+{\widetilde\Gamma})\lambda-\lambda^2-\frac{{\widetilde\Gamma}^2}{4}}}
    {\lambda}
=\frac{1}{\pi} \frac{\sqrt{(\lambda_+-\lambda)(\lambda-\lambda_- )}}{\lambda}\ , \!\!    &&
   \\
    \mbox{\rm with}\quad \lambda_\pm=\frac{\widetilde\Gamma}{2}+1\pm\sqrt{{\widetilde\Gamma}+1} \ . \!\!&&  \nonumber
\end{eqnarray}
This matches the expression~(\ref{eq:marchenko-pastur}) for the Marchenko-Pastur distribution. It also matches the leading density~(\ref{eq:leading-density-dyson-approach}) derived from eigenvalue repulsion in the Dyson gas approach of the previous subsection. An example is shown in figure~\ref{fig:MP-density}.

More generally, the generic structure will be the same for a more general class of potentials $V(\lambda)$  that give a  $\Gamma{=}0$ distribution with a left endpoint at $\lambda{=}0$. The functions $\Omega(X)$ and ${\widetilde\Omega}(X)$ will be of a more complicated form, but their values at $X=1$, (denoted $\Omega_c$ and ${\widetilde\Omega}_c)$ will be such that $S_c=2\sqrt{R_c}$, so that  $\lambda_-=S_c-2\sqrt{R_c}=0$. Using \eqref{eqn:eigenvalue_endpoints} and \eqref{eqn:largeN_String_eqn}, turning on $\widetilde\Gamma$ will then yield a gap from zero to the value:
\begin{equation}
    \lambda_-=\frac{{\widetilde\Omega}_c}{\Omega_c}\left\{\frac{\widetilde\Gamma}{\widetilde\Omega}_c+2-2\sqrt{\frac{\widetilde\Gamma}{\widetilde\Omega}_c+1}\right\}\ .
\end{equation}
For the class of models of interest for gravity, the potential will be tuned to get the multicritical behaviour of refs.~\cite{Morris:1990bw,Dalley:1992qg} at the left end. This is where $\lambda=S_c-2\sqrt{R_c}$. The ``double scaling limit" is the process of tuning to extract universal physics from the neighbourhood of this critical endpoint, and we shall do this next, {\it focusing on the issue of seeing a gap develop}.

\section{Gaps in the Double Scaling Limit and Gravity}

\label{sec:double-scaling-limit}

There is an order of limits issue here. The double scaling limit~\cite{Brezin:1990rb,Douglas:1990ve,Gross:1990vs,Gross:1990aw} zooms into the critical behavior at an endpoint, and all physics details away from that endpoint scale infinitely far away. So starting with a finite gap before the DSL process will only give the physics in the neighborhood of one or other edge of the gap, but the gap itself will become of infinite size and either become all that is left, or entirely disappear from the problem. Since we are interested in finite gaps within a given gravity spectrum (obtained using the DSL) it is clear that only gaps that are finite after the DSL are of interest, which means they are infinitesimal before the DSL.
In other words, we can focus our attention on scaling in to the endpoint $\lambda=S_c-2\sqrt{R_c}$ and use techniques that capture not 
only leading order at large $N$, and not only perturbative corrections in a $1/N$ expansion, but the complete non-perturbative physics. This is what the orthogonal polynomial methods of the last section allow. A simple example  is highly illustrative, and it is useful to study the simplest of the multicritical generalizations  alluded to in point {\bf (A)} of section~\ref{sec:wishart-and-more} (above equation~(\ref{eq:multicritical})).

\subsection{The Simplest Multicritical Model}

 The potential 
\be
\label{eq:example-critical-potential}
V(\lambda)=-4\lambda+2\lambda^2\ ,
\ee
results in a spectral density with an extra zero at the origin (changing it from the square root divergence of basic Wishart model to a square root fall off) as well as the usual square root at $\lambda=2$. (This is the case $k=1$ in equation~(\ref{eq:multicritical}).) 

\subsubsection{Leading Large $N$ Analysis}
This assertion can be checked explicitly by working out from equations~(\ref{eq:the-omegas}) that:
\begin{equation}
    \label{eq:the-omegas-11}
    {\widetilde\Omega}(X)=X-4R(X) \ , \quad\Omega=4(S(X)-1)\ ,
\end{equation}
and substituting into the equations~(\ref{eq:full-string-equations-discrete}) yield at leading large $N$ several solutions for $S(X)$ and $R(X)$ (recall that $X=n/N$). The non-trivial one of interest is:
\begin{equation}
    S(X) = \frac12+\frac16\sqrt{5+12X-4\sqrt{1+3X}}\ , \quad
    R(X) =\frac{X}{12}+\frac{1}{18}+\frac{1}{18}\sqrt{1+3X}\ .
\end{equation}
Note that for the range of $X$ relevant here, there is an equivalent functional form for $S(X)$ that is significantly simpler:\footnote{This result follows by mirroring some venerable techniques in elementary arithmetic  (see {\it e.g.}, ref.~\cite{chrystal1886algebra}, \S 9) that rely on that the fact that, for suitable $C$ and $D$, if one  writes $\sqrt{C-\sqrt{D}} = \sqrt{A}-\sqrt{B}$, then $\sqrt{C+\sqrt{D}} = \sqrt{A}+\sqrt{B}$ also holds. Multiplying them together gives $A+B$ while squaring either relation gives $A-B$.}
\be
S(X) = \frac{1}{3} \lr{1+ \sqrt{1+3X}}\ .
\ee
%
\begin{figure}
    \centering
\includegraphics[width=0.42\textwidth]{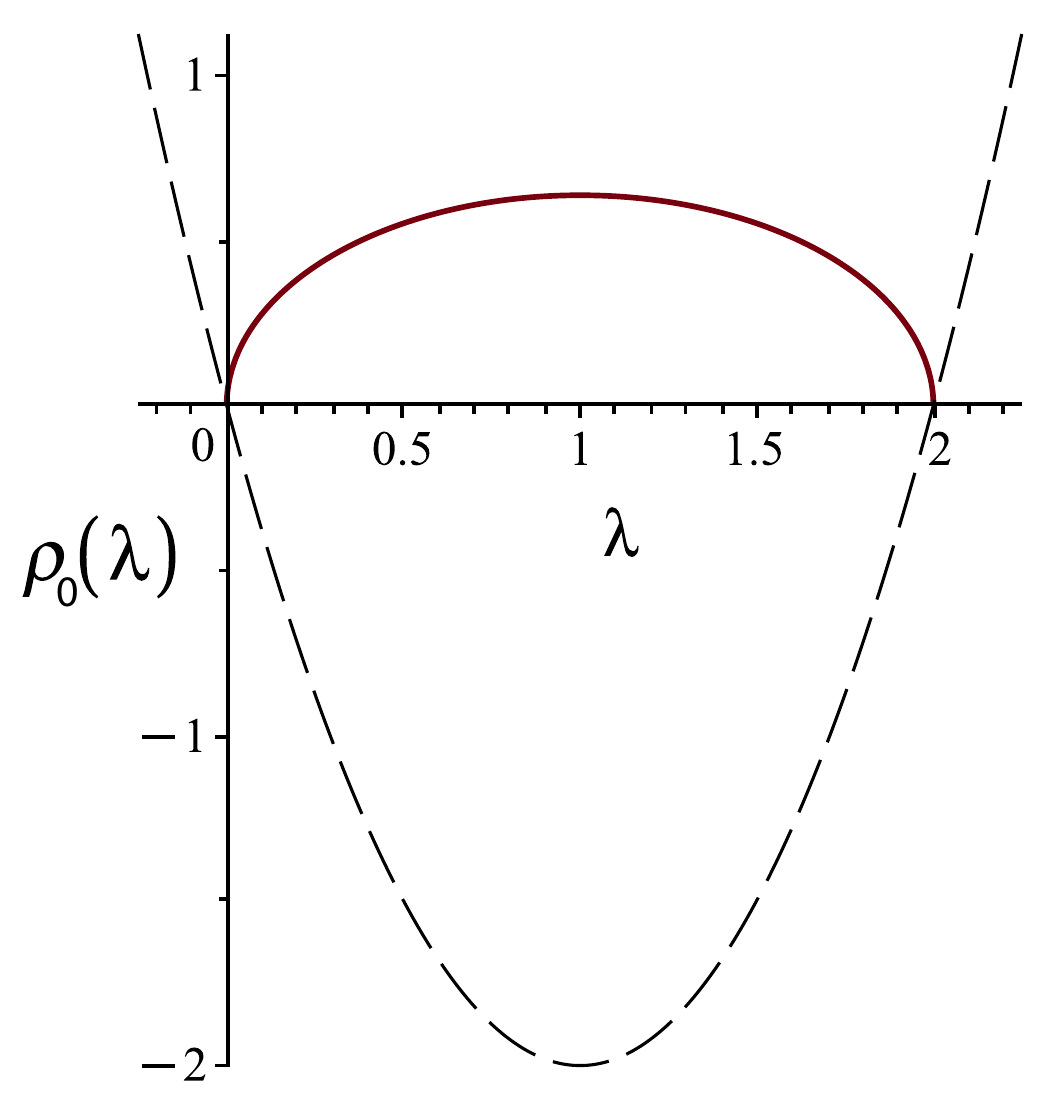}
    \caption{The density~(\ref{eq:1-1-density}) and potential (dashed line) for the simplest multicritical model. The endpoint at $\l=0$ has a square root edge $\rho_0 \sim \sqrt{\l}$ as opposed to a root divergence as in the Wishart case.}
    \label{fig:1-1-density-and-potential}
\end{figure}
Finally, using these in the 
representation~(\ref{eq:integral-representation}) gives:
\begin{equation}
\label{eq:1-1-density}
    \rho_0(\lambda)=\frac{2}{\pi}\sqrt{\lambda(2-\lambda)}\ ,
\end{equation}
as advertised. In fact it is a semi-circle law, centered on $\lambda=1$ (see figure~\ref{fig:1-1-density-and-potential}), but its non-perturbative physics is {\it not} that of an Hermitian matrix model, as will become clear.

\subsubsection{The Double Scaling Limit} 
In preparation for the double scaling limit, it is useful to be slightly more general than we have been so far. The potential~(\ref{eq:example-critical-potential})  is  precisely in its  critical form, meaning that it is  designed to produce the required extra zero at the endpoint. The matrix model can be defined with a potential that is slightly off-critical, with a parameter that must be tuned to arrive at criticality. There are many equivalent ways to do this, but the simplest way is to rescale $NV(\lambda)$ in the action to $\frac{N}{\gamma} V(\lambda)$, where the critical value of $\gamma$ is $\gamma_c=1$. The string equations~(\ref{eq:full-string-equations-large-N}) change only in that $\widetilde\Omega_n$  now has $\gamma X$ as its first term instead of $X$ (see equation~(\ref{eq:the-omegas})).

Notice that in the $k=1$ example, again we have $S_c\equiv S(1)=1$ and $R_c\equiv R(1)=\frac14$, and so the double scaling limit can be explored by defining the scaled variables: 
\begin{eqnarray}
    \label{eq:double-scaling}
   && \gamma=1+2\mu\delta^2\ , \quad X=1+2(x-\mu)\delta^{2}\ , \quad \gamma X=1+2x\delta^2\ ,\quad
    \frac{1}{N}\equiv\epsilon = 2\sqrt{2}\hbar\delta^3\ ,
    \\
    && S(X)=1-s(x)\delta^2\ ,\quad 
    R(X)=\frac14(1-2r(x)\delta^2)\ ,
    \nonumber
\end{eqnarray}
(where the numerical factors have been chosen to match conventions to follow), and $\delta$ will be taken to zero in the limit. The precise combination of powers of $\delta$ are such that non-trivial physics survives the limit (see below). The coordinate $x$ can range over the whole real line, and indeed finding sensible solutions for $r(x)$ and $s(x)$ over that full support is equivalent to finding the full family of orthogonal polynomials from which all the (double-scaled) physics can be constructed. Note that   $X=1$ coincides with $x{=}\mu$. For the examples that will follow, $\mu$ will turn out to be positive.

Since the endpoints are at $\lambda_{\pm}=S_c\pm2\sqrt{R_c}$, it is the scaling variables $w(x)=r(x)+s(x)$  and $u(x)=r(x)-s(x)$  that will describe the scaled physics there, and all physics factorizes into these two sectors at criticality~\cite{Dalley:1992qg,Dalley:1991jp}. Our interest is in the $\lambda_-=0$ end, and hence the  function $u(x)$. Substituting the scaling ans\"atze~(\ref{eq:double-scaling}) into~(\ref{eq:the-omegas-11}) (remembering that now $\widetilde\Omega = \gamma X -4R(X)$) and Taylor expanding, gives for the first few orders in $\delta$ (temporarily dropping the indication of  $x$ dependence on $u$ and $w$ for clarity): 
\begin{eqnarray}
 {\widetilde\Omega}(X+\epsilon)&=&  
 [u+w+2x]\delta^2
 +\sqrt{2}\hbar[2+w^\prime+u^\prime] \delta^3
 +\hbar^2 [u^{\prime\prime}+w^{\prime\prime}] \delta^4\cdots \nonumber\\
 \Omega(X-\epsilon)&=& 
 2[u-w]\delta^2
 +2\sqrt{2}\hbar[w^\prime-u^\prime]\delta^3
 +2\hbar^2 [u^{\prime\prime}-w^{\prime\prime}] \delta^4\cdots 
\end{eqnarray}
whereupon substitution of these and expressions~(\ref{eq:the-omegas-11}) into the main equations~(\ref{eq:full-string-equations-large-N}) gives:
\begin{eqnarray}
 0&=&  
 4(w+x)\delta^2
 +\sqrt{2}\hbar(2+w^\prime+u^\prime+2\Gamma) \delta^3
 + [\hbar^2u^{\prime\prime}+\hbar^2w^{\prime\prime}-(u-w)^2] \delta^4\cdots \nonumber\\
 0&=& 
 -4(u+x)(w+x)\delta^4
 -\sqrt{2}\hbar[w(w^\prime-u^\prime+2\Gamma)+u(u^\prime-w^\prime+2\Gamma) +4\Gamma x]\delta^5
 \nonumber \\&& \hskip7cm +(u-w) [(\hbar^2u^{\prime\prime}-u^2)- (\hbar^2w^{\prime\prime}  - w^2)]  \delta^6\cdots 
\end{eqnarray}
Solving the first for $w(x)$ gives: 
\begin{equation}
    w(x) = -x -\frac{\sqrt{2}}{4}\hbar\left[u^\prime(x)+2\Gamma+1\right]\delta+
\frac18\left[2(u(x)+x)^2-\hbar^2u^{\prime\prime}\right]\delta^2
    +\cdots
\end{equation}
and substituting it into the second yields, at order $\delta^6$, the following non-trivial equation for $u(x)$:
\begin{equation}
    \label{eq:big-string-equation}
    u{\cal R}^2-\frac{\hbar^2}{2}{\cal R}{\cal R}^{\prime\prime}+\frac{\hbar^2}{4}({\cal R}^\prime)^2 = \hbar^2\Gamma^2\ ,
\end{equation}
where here ${\cal R}\equiv u+x$. (Dividing through by $\delta^6$ and sending $\delta\to0$ sends all other terms in the expansion to zero, showing that this equation  fully defines the physics of the double-scaled matrix model non-perturbatively. This form of string equation, first found and explored in refs.~\cite{Morris:1990bw,Dalley:1991qg} (several higher order examples are systematically presented in detail in the latter), with the generalization to include~$\Gamma$ found in ref.~\cite{Dalley:1992br}, has been explored a lot in the context of dilaton gravity in recent times, and will attract our attention for the study of gaps shortly. The more general case discussed in subsection~\ref{sec:multicritical} will continue to have this overall structure.

\subsubsection{Double-Scaling the Leading Spectral Density}
Before studying the string equation closely, it is useful to double scale the integral representation~(\ref{eq:integral-representation}) itself. which will lead to a useful counterpart for double scaled quantities. Substituting the scaling~(\ref{eq:double-scaling}) into the integral representation~(\ref{eq:integral-representation}), along with $\lambda = 0+E\delta^2$ yields a scaled spectral density:
\begin{equation}
\label{eq:scaled-integral-represention}
    \rho_0(E) = \frac{1}{2\pi\hbar}\int_{-\infty}^\mu\frac{\Theta(E-u_0(x))}{\sqrt{E-u_0(x)}}dx\ .
\end{equation}
where the density was multiplied by a factor of $N$ to make it finite in the scaling limit, and a factor of $1/\sqrt{2}$ has been included to match conventions in the literature to be discussed. The notation $u_0(x)$ indicates we solve \eqref{eq:big-string-equation} to leading order in the $\hbar$ expansion, which gives the leading order solution to the density of states. 


Continuing the discussion of the $k{=}1$ example of the previous subsection, there are two important pieces of leading behavior for solutions of the equation~(\ref{eq:big-string-equation}):
\begin{eqnarray}
    \label{eq:asymptotics}
    &&u(x)=-x+\frac{\hbar\Gamma}{(-x)^\frac12}-\frac{\hbar^2\Gamma^2}{2x^2}\cdots\ ,\qquad\mbox{\rm as}\quad x\to-\infty\nonumber\\
    &&u(x)=0+\frac{\hbar^2\left(\Gamma^2-\frac14\right)}{x^2}+\cdots\ , \qquad\mbox{\rm as}\quad x\to+\infty\ . 
\end{eqnarray}
The leading portion of the negative $x$ 
 solution translates to $\rho_0(E)\sim E^{\frac12}/\pi\hbar$, as  expected for  the edge of semi-circle distribution. Setting $\Gamma{=}0$ for now, it can be seen that $u(x)=-x$ is an exact solution to the string equation to all orders in the $x\to-\infty$ expansion. This is not the Airy model however, which would have $u(x)=-x$ for positive $x$ as well. Instead, with the choice $\mu=0$ (so that there is no perturbative contribution from the postive $x$ sector), all the physics is perturbatively the same as Airy, but differs non-perturbaively due to the fact that the spectrum must be positive. This non-perturbative departure is the part of the role of the positive $x$ sector. If $\mu>0$, so that the integral runs into the positive $x$ sector, there is new  physics to take into account.

\subsection{The Bessel Models}
\label{sec:bessel-models}
 Consider  the case of $\mu>0$. For now it can be chosen as $\mu=1$ for the sake of illustration. Now, something new can occur in perturbation theory. Looking at the leading $x>0$ solution in equation~(\ref{eq:asymptotics}),  $u_0(x)=0$, the resulting contribution to the spectral density from doing the integral~(\ref{eq:scaled-integral-represention}) in that regime is $\rho_0(E)=\mu/2\pi\sqrt{E}$, the square root divergence familiar from the simple Wishart hard wall prototype we studied in section~\ref{sec:wishart-and-more}.

 The next order in equation~(\ref{eq:asymptotics}), with $u(x){=}\hbar^2(\Gamma^2-\frac14)/x^2$, has certain special features that are present in all multicritical solutions of the equation that have a $u{=}0$ leading asymptote. Let us consider  this form of $u(x)$ as a solution in its own right, ignoring the $x<0$ sector for now. First noticed in refs.~\cite{Dalley:1992qg} as a puzzling solution (dubbed ``topological'' there), it has since been recognized~\cite{Carlisle:2005wa,Johnson:2020heh} as characterizing the ``Bessel model'', which is in fact a scaling limit of the discrete Wishart model of section~\ref{sec:wishart-and-more}. The square root divergence seen above is the leading perturbative part, and then there is an infinite family of perturbative and non-perturbative completions characterized by $\Gamma$, built from
 Bessel functions
 (essentially the scaling limit of the  orthogonal polynomials).

\begin{figure}
    \centering
\includegraphics[width=0.75\textwidth]{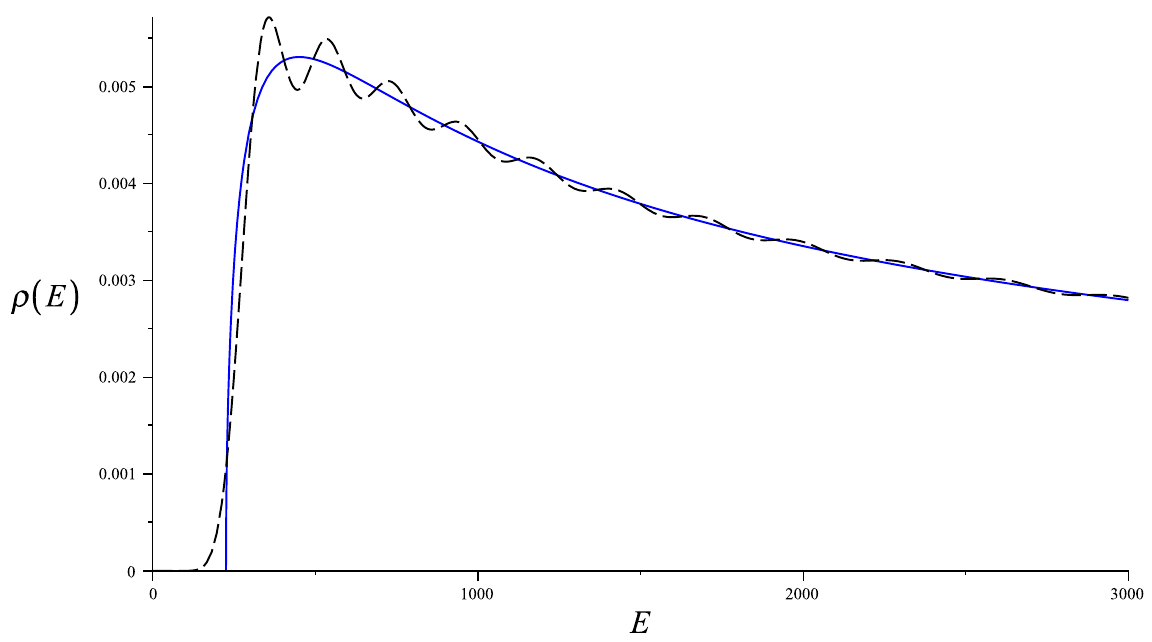}
    \caption{\rm The leading form of the gap in the Bessel model spectrum is shown as a solid line.  Here, $\hbar=1$ and ${\widetilde\Gamma}=15$, and the gap is at $E{=}{\widetilde\Gamma}^2=225$. The non-perturbative spectrum, superimposed as a dashed line, includes corrections that soften the gap.}
    \label{fig:bessel-gap}
\end{figure}

The complete spectral density for these models can be written in closed form as~\cite{doi:10.1063/1.530157}:
\begin{equation}
\label{fig:bessel-density-exact}
    \rho(E)=\frac{1}{4\hbar^2}\left[J_\Gamma^2(\xi)+J_{\Gamma+1}^2-\frac{2\Gamma}{\xi}J_\Gamma(\xi)J_{\Gamma+1}(\xi)\right]\ .
\end{equation}
See refs.~\cite{Johnson:2020heh,Johnson:2021rsh} for a  derivation and a recent discussion in the context of the  ${\cal N}=1$ JT supergravity quenched free energy.
It  was noticed in ref.\cite{Johnson:2021rsh} that the spectrum has an effective gap that grows with~$\Gamma$, and this feature was suggested as a potential model of black hole gaps.  

There is a straightforward way~\cite{Johnson:2023ofr} of studying the gap for these models in the large $\Gamma$ limit  and it will be a prototype for what will come in later sections when we discuss   JT supergravity and black holes. One simply follows the logic of previous subsections, imagining that $\Gamma$ is a significant fraction of~$N$, writing it as $\Gamma=N{\widetilde\Gamma}$, and subsequently incorporating $\widetilde\Gamma$ into the leading large $N$ analysis.  However, now  that we are in the double scaling limit, since $\hbar$ is the renormalized $1/N$, the condition to explore for finite gaps  is $\hbar\Gamma={\tilde \Gamma}$. This gives us an effective classical solution $u_0(x)={\widetilde\Gamma}^2/x^2$, which can be inserted into the scaled integral representation~(\ref{eq:scaled-integral-represention}) (with $\mu{=}1$) to give a remarkably simple form:
\begin{equation}
\label{eq:bessel-gap-density}
    \rho_0(E) = \frac{1}{2\pi\hbar}\frac{\sqrt{E-{\widetilde\Gamma}^2}}{E}\ , \qquad\mbox{\rm (large order Bessel model)\ .}
\end{equation}
 The spectrum starts,  after a gap, at $E={\widetilde\Gamma}^2$ and at large $E$ it falls off as $1/2\pi\sqrt{E}$.  Figure~\ref{fig:bessel-gap} shows an example, with the exact Bessel density~(\ref{fig:bessel-density-exact}) overlaid for comparison. 
The latter includes non-perturbative corrections that soften the sharpness of the gap.

The scaling of the gap with $\widetilde\Gamma$ should be contrasted with with the value, $\lambda_-$, of  the Marchenko-Pastur spectrum, computed before double scaling (see equation~(\ref{eq:unscaled-gaps}). It is linear with a square root correction. Note however that for small ${\widetilde\Gamma}=\Gamma/N$ as defined there, the scaling of the gap also becomes quadratic at leading order,
 $\lambda_-=\frac18{\widetilde\Gamma}^2+\cdots$, perhaps fitting with the fact that in the scaling limit, the entire spectrum is only a small part of the whole (unscaled) matrix model's spectrum.

\subsection{Double-Scaling Higher Multicritical  Models}
\label{sec:multicritical}
As anticipated in section~\ref{sec:double-scaling-limit} the entire family of multicritical models will be needed, as they will be building blocks for matrix models of various kinds of JT supergravity. They are indexed by $k$, giving that number of zeros at the origin for the (unscaled) critical leading order density~(\ref{eq:multicritical}). The critical potentials that yield those densities are\cite{Dalley:1992qg}:

\begin{equation}
\label{eq:critical-potentials}
    V_k(\lambda) = \sum_{p=1}^{k+1}\frac{\left(-1\right)^{p +1} \left(k +1\right)! \Gamma \! \left(\frac{1}{2}-k \right)}{2^{p -1} p \left(k -p +1\right)! \Gamma \! \left(\frac{1}{2}-k +p \right)}\lambda^p\ .
\end{equation}

\begin{figure}
    \centering
    \includegraphics[width=0.42\textwidth]{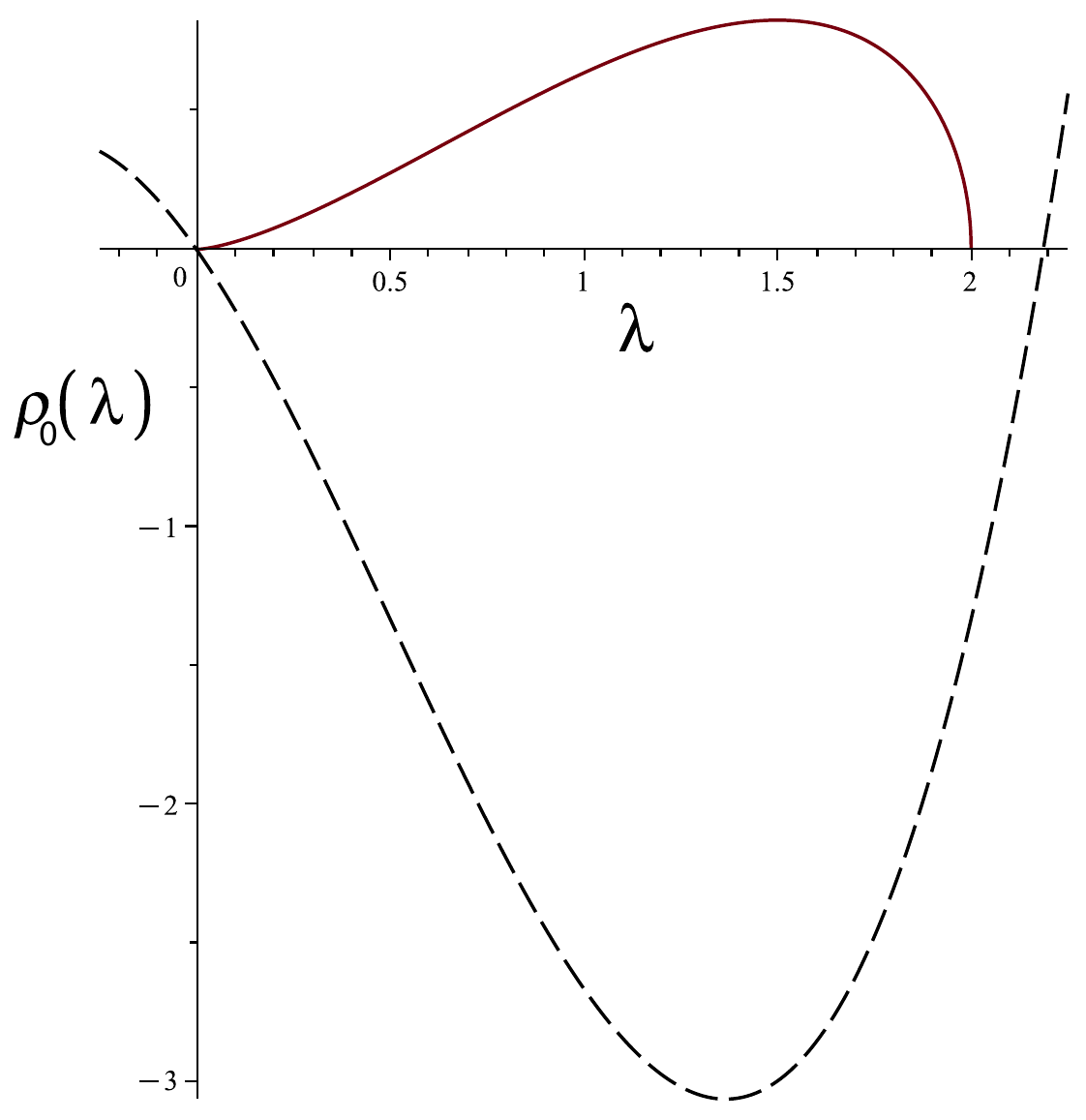}
  \caption{The $k=2$ multicritical density of states (red) given by equation~\eqref{eq:multicritical} and corresponding potential $V(\l)$ (dashed line). The multicritical densities have a scaling $\rho_0 \sim \l^{k-\frac{1}{2}}$ near the edge of the spectrum around $\l=0$. The other edge has the standard root scaling.} 
\label{fig:2-1-density-and-potential}
\end{figure}
The $k=2$ example is illustrated in figure~\ref{fig:2-1-density-and-potential}.
The ans\"atze:
\begin{eqnarray}
    \label{eq:double-scalingk}
   &&
   X=1+2x\delta^{2k}\ ,\quad
    \frac{1}{N}\equiv\epsilon = 2\sqrt{2}\hbar\delta^{2k+1}\ ,
    \\
    && 
    S(X)=1-s(x)\delta^2\ ,\quad 
    R(X)=\frac14(1-2r(x)\delta^2)\ ,
    \nonumber
\end{eqnarray}
put into the defining matrix model equations~(\ref{eq:full-string-equations-discrete}) yield at order $\delta^{2k+4}$ the string equation of the form~(\ref{eq:big-string-equation}) where now ${\cal R} \equiv R_k[u]+x$. The $R_k[u]$ are the ``Gel'fand-Dikii''~\cite{Gelfand:1975rn} differential polynomials  in $u(x)$ and its derivatives, in a normalization such that the non-derivative part has unit coefficient: $R_k{=}u^k+\cdots+\#u^{(2k-2)}$ where $u^{(m)}$ means the $m$th $x$-derivative.\footnote{For example: 
\be
\label{eq:gelfand-dikii-R}
R_1{=}u \ ,\qquad  R_2{=}u^2{-}\frac{\hbar^2}{3}u^{\prime\prime}\,, \qquad R_3 {=} u^3+\frac{\hbar^2}{2}(u^\prime)^2+{\hbar^2}uu^{\prime\prime}+\frac{\hbar^4}{10}u^{\prime\prime\prime\prime}\,.
\ee
}
Successive $R_k[u]$ can be obtained using a recursion relation, which we won't need here. More generally, since the defining  matrix model equations~(\ref{eq:full-string-equations-discrete}) are linear in the matrix model potential $V(\lambda)$, it is possible to sum~\cite{Gross:1990vs} the critical potentials $V_k(\lambda)$ with coefficients~$t_k$ (with additional factors of powers of $\delta$ to adjust the scaling appropriately) so as to construct a more general model. The string equation is again of the form~(\ref{eq:big-string-equation}), but now with~\cite{Dalley:1992qg}:
\begin{equation}
    {\cal R}\equiv\sum_{t=1}^\infty t_k R_k[u]+x\ .
\end{equation}
A first pass at the form of $u_0(x)$, the leading part of $u(x)$, can be obtained by dropping all derivatives (and hence all powers of $\hbar$), in which case $R_k[u]$ becomes $u_0^k$. In preparation for what is to come, hold fixed $\widetilde\Gamma{=}\hbar\Gamma$ as done before,   and then the string equation~(\ref{eq:big-string-equation}) becomes:
\begin{equation}
\label{eq:leading-special}
    u_0{\cal R}_0^2={\widetilde\Gamma}^2\ ,
\end{equation}
where ${\cal R}_0\equiv\sum_{k=1}^\infty t_k u_0^k+x$. In the case of vanishing $\widetilde\Gamma$, there are two obvious solutions of the equation, either ${\cal R}_0=0$ or $u_0=0$. The former will be taken for the $x<0$ regime, and the latter for the $x>0$ regime. Indeed, the leading behaviour of the $k=1$ example~(\ref{eq:asymptotics}) was such an example pair of these solutions. For any of the $k$th models (set that $t_k=1$) and all the others to zero in ${\cal R}_0$), this is the form of the solution.

The linear combination of models will be of particular interest moving forward. Beginning with the work of refs.~\cite{Johnson:2020heh,Johnson:2020exp} it was shown that specific values for the $t_k$ can be found that allow us to build a matrix model for various ${\cal N}{=}1$ JT supergravity theories of interest. This will be reviewed in section~\ref{sec:Neq1-JT-supergravity}. For the case of ${\cal N}{=}2$ it was realized in ref.~\cite{Johnson:2023ofr} that the ansatz $u_0=0$ for $x>0$ would not suffice for reproducing the  form of the spectral density. This is clear from equation~(\ref{eq:leading-special}), where with $\widetilde\Gamma$ non-zero, a new asymptotic behaviour becomes apparent if one assumes that $u_0(x)$ is small compared to $x$. In that case, ${\cal R}_0$ becomes essentially $x$, and then the solution for $x>0$ is $u_0={\widetilde\Gamma}^2/x^2$. How this new ansatz is used to construct the ${\cal N}{=}2$ JT supergravity from multicritical models is reviewed in section~\ref{sec:Neq2-JT-supergravity}, and it will be used in section~\ref{sec:Neq4-JT-supergravity} to construct ${\cal N}{=}4$ JT supergravity.

\section{JT Supergravity}
\label{sec:JT-gravity}
The previous section has introduced all the building blocks needed to study two dimensional supergravity theories that, while worthy of study in their own right as model dilaton-gravity systems,  also can appear  as a crucial part of the near-horizon dynamics of higher dimensional systems containing black holes.  As already explained, it is the latter aspect that that is our main motivation.

The key point is that all the known supersymmetric  extensions of JT gravity can be fully non-perturbatively constructed as random matrix models by using the multicritical building blocks described. As already explained, the multicritical models have a natural and robust mechanism for  generating a gap in the leading order supergravity spectrum when there is a large $E{=}0$ degeneracy present. 

It seems likely that one can always expect there to be {\it some kind} of effective  random matrix model description of the effective throat geometry near the horizon of a black hole, the degeneracy will always result, through eigenvalue repulsion, a gap in the spectrum of the kind seen in the examples given here.

In the following subsections, the multicritical model construction of various JT supergravity models  is described, beginning with a review of what was done in refs.~\cite{Johnson:2020heh,Johnson:2020exp} for ${\cal N}{=}1$. While that case does not have traditional BPS states {\it per se}, the role of degeneracy is played by the ``Ramond punctures" discussed in this context in ref.~\cite{Stanford:2019vob}, and indeed a factor ${\rm e}^{-S_0}\Gamma=\hbar\Gamma$ results from such an insertion. We will not say much about  large $\Gamma$ in this case, however, moving swiftly on to the cases with extended supersymmetry. 

There,
$\Gamma$ readily has an interpretation in terms of counting BPS states of black holes. The method for fully non-perturbatively defining the matrix model of ${\cal N}{=}2$ JT supergravity out of the multicritical building blocks was recently discovered in ref.~\cite{Johnson:2023ofr}, (following the initial casting of it as a random matrix model to all orders in perturbation theory in ref.~\cite{Turiaci:2023jfa}). The construction will be reviewed in  subsection~\ref{sec:Neq2-JT-supergravity}. The multicritical construction  of  the matrix model of ${\cal N}{=}4$ JT supergravity  
will  be presented here for the first time in subsection~\ref{sec:Neq4-JT-supergravity} following.

\subsection{$\mathcal{N}=1$ JT Supergravity}
\label{sec:Neq1-JT-supergravity}
As warmup for more complicated models, let us  begin by discussing $\mathcal{N}=1$ JT gravity and its corresponding random matrix model dual \cite{Stanford:2019vob}. The $\mathcal{N}=1$ quantum mechanical system dual to $\mathcal{N}=1$ JT has Hamiltonian given by $H = \mathcal{Q}^2$ where $\mathcal{Q}$ is the Hermitian supercharge. There is an operator $(-1)^F$ that distinguishes bosonic from fermionic states. In a basis where $(-1)^F = \t{diag}\lr{I,-I}$ the supercharge takes the form
\be
\mathcal{Q} = \begin{pmatrix}
0 & Q \\
Q^\dag & 0 
\end{pmatrix} \,,
\ee
where $Q$ is an $N_b \times N_f$ complex matrix introduced in \eqref{eq:probability} giving a random Hamiltonian with $N_{b}$ bosonic and $ N_f$ fermionic states. The standard $\mathcal{N}=1$ super JT theory has the same number of bosonic and fermionic states and has no ground states with $E=0$. Ref.~\cite{Stanford:2019vob} showed perturbatively (in the topological expansion) that this JT supergravity theory is also described as the double scaling limit of random matrices $Q$ with $N_b = N_f=N$ with a finely tuned potential, which we now explain. The disc order partition function for ${\cal N}=1$ JT supergravity is 
\begin{equation}
    Z(\beta) = {\rm e}^{S_0}\frac{1}{2\sqrt{\pi\beta}}{\rm e}^{\frac{\pi^2}{\beta}}\ ,
\end{equation}
where $S_0$ is the extremal ($1/\beta{\equiv}T{=}0$) entropy, a parameter that multiplies the (topological in 2D) Einstein-Hilbert term in the action.
The partition function is the Laplace transform of the spectral density:
\begin{equation}
\label{eq:leading-spectral-Neq1}
    \rho_0(E) = \frac{\cosh\bigl(2\pi\sqrt{E}\bigr)}{2\pi\hbar\sqrt{E}}\ ,
\end{equation}
where we have written $\hbar\equiv {\rm e}^{-S_0}$, {\it i.e.,} the  gravity theory's  topological expansion parameter is given the same symbol, $\hbar$, used earlier~(see eqns.~(\ref{eq:double-scaling}) and~(\ref{eq:double-scalingk})) for  the (renormalized $1/N$) matrix model topological expansion parameter, since they are to be identified. 
Ref.~\cite{Stanford:2019vob} then proceeds by using the leading spectral density~(\ref{eq:leading-spectral-Neq1}) as a seed for defining the matrix model perturbatively by recursion relations derived from the loop equations of the model, a procedure that generalizes to super-Riemann surfaces the topological recursion~\cite{Mirzakhani:2006fta,Eynard:2007fi,Eynard:2014zxa} used for defining the ordinary JT case of ref.~\cite{Saad:2019lba}.

To define the  random matrix model fully non-perturbatively (meaning that it has the perturbation theory  as an asymptotic expansion, but  non-perturbative physics besides that), a route to follow~\cite{Johnson:2020heh} is the method of  building the model out of the  multicritical model building blocks of the last section. This procedure works as follows.

The spectral density can be expanded in powers of $E$ as follows:
\begin{equation}
\label{eq:expansion1}
    \rho_0(E) = \frac{1}{2\pi\hbar\sqrt{E}}+\frac{1}{2\pi\hbar}\sum_{k=1}^\infty \frac{(2\pi)^{2k}}{(2k)!} E^{k-\frac12}\ .
\end{equation}
 The positive powers of $E$ appearing in the sum are familiar from  multicritical random matrix models with some double scaled string equation of the form $u_0(x)^k{+}x=0$. So at this order, a linear combination  of the kind discussed in   section~\ref{sec:multicritical}, $\sum_k t_k u_0^k+x=0$ should be able to reproduce this leading density (through the integral transform~(\ref{eq:scaled-integral-represention})), for an appropriate choice of $t_k$.  
Indeed, it is useful to rewrite the integral representation  as a $u_0$ integral, where the Jacobian is $\partial x/\partial u_0{=}{-}\sum_k k t_k u_0^{k-1}$, and hence, integrating up to some $x=\mu$:
\be
\label{eq:u-integral}
\rho_0(E) = \frac{1}{2\pi \hbar} \sum_{k=1}^\infty  \int_{E_0}^E \frac{ k t_k u_0^{k-1}}{\sqrt{E-u_0}} d u_0 \ ,\qquad \mbox{\rm (partial contribution)}
\ee
where the energy $E_0{=}u_0(\mu)$. For the current case, we should integrate this form of $u_0(x)$ right up to $x{=}0$ and hence $\mu{=}0$, and there $u_0{=}0$ and so $E_0$ vanishes. The cases to come will requite different choices. For the current case, the integral can be straightforwardly carried out with the result: 
\be
\label{eq:u-integral2}
\rho_0(E) = \frac{1}{2\pi \hbar} \sum_{k=1}^\infty  t_k 
 \frac{2^{2k{-}1}k ((k{-}1)!)^2}{(2k{-}1)!} E^{k-\frac12} \ ,\qquad \mbox{\rm (partial contribution)}
\ee

What makes the construction stronger and particular to the supersymmetric context is  the $E^{-\frac12}$ tail as well, which  as we've seen comes from double scaling multicritical models of {\it positive} Hamiltonians, which are natural for supersymmetry. As discussed in section~\ref{sec:bessel-models}, this  behaviour of the tail comes from integrating over the $x>0$ region where the leading behaviour is $u_0(x)=0$ after putting this into the representation~(\ref{eq:scaled-integral-represention}). All that is needed is to match to the problem in question to determine the value of upper limit $\mu$. For this example, upon inspection the results are:
\begin{equation}
\label{eq:teekay1}
    t_k=\frac{\pi^{2k}}{(k!)^2}\ .\qquad \mu = 1 \ .
\end{equation}
The leading  string equation 
${\cal R}_0\equiv \sum_{k=1}^\infty t_k u_0^k+x=0$   resulting from this can be written  as:
\begin{equation}
\label{eq:leading-u0-1}
    I_0(2\pi\sqrt{u_0}) - 1 +x = 0\ .
\end{equation}
The full string equation~(\ref{eq:big-string-equation}) is  solved with leading asymptotic behaviour for $u_0(x)$ determined by $u_0(x)=0$ for positive $x$ and equation~(\ref{eq:leading-u0-1}) for $x<0$, with  a smooth solution that interpolates between these two regimes in the interior~\cite{Johnson:2020exp}. The precise nature of the solution depends upon $\Gamma$, and the cases $\Gamma=0,\pm\frac12$ (corresponding to $\boldsymbol{\alpha}=1,0$ or $2$ in ref.~\cite{Stanford:2019vob}) featured prominently in refs.~\cite{Stanford:2019vob,Johnson:2020heh,Johnson:2020exp}. 

More generally, we can consider modifying the class of models under consideration by introducing some number of supersymmetric ground states into the quantum mechanical boundary theory. From the random matrix model we would take $Q$ to be a $\lr{\Gamma + N} \times N$ complex random matrix,  giving, as we've seen, an  additional $\Gamma$ zero energy supersymmetric states. Incorporating these ground states into the $\mathcal{N}=1$ JT description can be accomplished by introducing additional insertions known as ``Ramond punctures''~\cite{Witten:2012ga} in the path integral when performing the sum over surfaces. (See section 5.5 of ref.~\cite{Stanford:2019vob} for more discussion of this, where they use the symbol $\nu$ instead of~$\Gamma$.) A puncture is assigned, by hand, a weight $\Gamma {\rm e}^{-S_0}$, and gives an additional contribution of $\Gamma$ when computing the disk amplitude. Here,  it is interesting to consider $\Gamma$  large, as it is already apparent~\cite{Johnson:2020heh} that the solutions for $u(x)$ and the resulting spectral density, are such that a gap begins to develop in the spectral density. 
However, we will not pursue this here, moving on instead to the cases with extended supersymmetry.

\subsection{$\mathcal{N}=2$ JT Supergravity}
\label{sec:Neq2-JT-supergravity}

We now consider the random matrix model dual to $\mathcal{N}=2$ JT supergravity, as first discussed perturbatively (in topology) in ref.~\cite{Turiaci:2023jfa}, and non-perturbatively constructed in ref.~\cite{Johnson:2023ofr}.
This model of gravity arises naturally in considering  black holes dual to near-$\frac{1}{16}$th-BPS states in ${\cal N}{=}4$ supersymmetric Yang-Mills~\cite{Boruch:2022tno}, as reviewed in Appendix~\ref{sec:gravity-embedding}.

In this case the Hamiltonian is given by $H {=} \{\mathcal{Q}, \mathcal{Q}^\dag \}$ where  $\mathcal{Q}$ is  the supercharge. There is a~$U(1)_R$ symmetry under which  $\mathcal Q$  has charge $\hat{q}$. The states organize into multiplets, with BPS states being annihilated by $\mathcal{Q}, \mathcal{Q}^\dag$ and having any possible integer R-charge $k$. The non-BPS states fall into multiplets of adjacent R-charges $(k,k+\hat{q})$. The Hilbert space decomposes into a sum over sectors of fixed charge $\mathcal{H} {=} \oplus_k \mathcal{H}_k$, where each sector can be decomposed as $\mathcal{H}_k = \mathcal{H}_k^{\t{BPS}} \oplus \mathcal{H}_k^{+} \oplus \mathcal{H}_k^{-}$. The BPS sector is self-explanatory, while the $\mathcal{H}_k^{+},\mathcal{H}_k^{-}$ sectors are states of charge $k$ that are in the multiplets $(k,k+\hat{q})$ and $(k-\hat{q},k)$ respectively.

It is not immediately obvious how to model such a system using random random matrices. The conclusion of \cite{Turiaci:2023jfa} is that adjacent supermultiplets $(k,k+\hat{q})$ and $(k-\hat{q},k)$ should be taken to be statistically independent. A given multiplet $(k,k+\hat{q})$ is described by an AZ ensemble with $(\alpha,\beta)=(1+2\Gamma,2)$ where $\Gamma$ is the number of BPS states at R-charge $k$ or $k+\hat{q}$.\footnote{Generically, there will only be a non-zero number of BPS states at charge $k$ or $k+\hat{q}$ but not both.} The full random matrix model is then given by modeling each supermultiplet by an independent random matrix. We now explain how the theory can be directly obtained in the double scaling limit.

The  partition function of the model can be expressed as a sum over sectors labeled by the $U(1)$ R-charge with multiplets labelled by a modified notation $(q-\frac{\hat{q}}{2},q+\frac{\hat{q}}{2})$. Within a sector, the leading spectral density of a non-BPS multiplet has  the following form:
\begin{equation}
\label{eq:spectral-density-Neq2}
    \rho_0(E) =   \frac{{\rm e}^{S_0}}{2\pi}\frac{\sinh\bigl(2\pi\sqrt{E-E_0(q)}\bigr)}{4\pi^2 E}\Theta\bigl(E-E_0(q)\bigr)\ ; \qquad E_0(q)=\frac{q^2}{{4\hat q}^2}\ ,
\end{equation}
where ${\hat q}$ is the  R-charge of the supercharge $Q$, and $S_0$ is  the extremal   black hole entropy. The BPS multiplet with charge $(q-\frac{\hat{q}}{2})$ has density:
\begin{equation}
\label{eq:BPS-density-Neq2}
   \rho_0^{\rm BPS}(E)=  \frac{{\rm e}^{S_0}}{4\pi^2}\sin\left(\frac{\pi q}{\hat q}\right)\, \delta(E) \ ,
\end{equation}
and (crucially) only exists for $q<{\hat q}$. Note that there a convention choice of energy scale $E_{\rm schw}=2$ has been made in the spectrum~(\ref{eq:spectral-density-Neq2}). The scale $E_{\rm schw}$ is determined by the details of the embedding of the effective super-Schwarzian (JT) theory into the full black hole supergravity geometry.

 In the semi-classical regime, $S_0$ is already a large number, and so after exponentiating there is an extremely large degeneracy of BPS states at $E=0$. There is also a finite gap, set by $E_0(q)$. 
 An {\it essential  point} here  is that it is key that the gap and the large degeneracy  come together. 
Ref.~\cite{Johnson:2023ofr} showed that  what's remarkable is that the above { specific} formulae, while arising from study of the intricate properties of the theory as an ${\cal N}{=}2$ supergravity theory,  are {\it required} to be of that form if there is a description of the matrix model as a multicritical model in the sense of last section. 

The result works as follows. Expanding the spectral density by first Taylor expanding the argument of the hyperbolic sine in $E_0/E$ and then expanding the hyperbolic sine, the result is of a form that generalizes~(\ref{eq:expansion1}):
\begin{equation}
\label{eq:expansion2}
    \rho_0(E) = \frac{1}{\hbar}\sum_{k=1}^\infty A_k(E_0) E^{k-\frac12}+ \frac{1}{\hbar}\sum_{k=2}^\infty B_k(E_0) E^{\frac12-k}+\frac{1}{\hbar}\frac{B_1}{\sqrt{E}}\ ,
\end{equation}
where here $B_1=1/4\pi^2$. For the case of $E_0{=}0$, all the higher $B_k=0$ and the model again has some expansion of the form seen in the ${\cal N}=1$ case. This shows that the matrix model  can be built from multicritcal models in a very similar way to that case, with $u_0$ solving equation~(\ref{eq:leading-special}) by having  ${\cal R}_0{=}0$ for $x{<}0$, and $u_0{=}0$ for $x{>}0$, and  this time~\cite{Johnson:2023ofr}:
\begin{equation}
\label{eq:teekay2}
    t_k=\frac{\pi^{2k-1}}{2(2k+1)(k!)^2}\ .\qquad \mu = \frac{1}{2\pi} \ .
\end{equation}

When $E_0{\neq}0$ the problem becomes much more complicated, since there is an infinite number of terms going to successively higher inverse powers of $E$. If there is some construction in terms of multicritical models, such terms could have at least two sources:

\begin{enumerate}[{\bf (1)}]
\item Equation~(\ref{eq:leading-special}) mut be solved with a $u_0(x)$ that yields $R_0{=}0$ now only asymptotically as $x{\to}{-}\infty$. There could be $E_0$ dependence in the required $t_k$ coefficients. Let us denote them ${\tilde t}_k(E_0)$. 
 Indeed, by matching positive powers of $E$, we can  deduce expressions for the ${\tilde t}_k(E_0)$, at least order by order in a small $E_0$ expansion. 
 
 \item The individual multicritical models combined in this way must be of a form such that  their continuum contribution to the spectrum begins at gap energy $E_0$. 
 Using equation~(\ref{eq:u-integral}), this  generates a specific pattern of negative powers of $E$, {\it e.g.,} from expanding the density of the $k$th model:
 \begin{eqnarray}
\label{eq:examples-of-minimals}
\rho_0^{(1)}\! &=& 
2\frac{(E-E_0)^\frac12}{2\pi\hbar}\ ,\,\, \rho_0^{(2)} = 
\frac83\frac{(E-E_0)^\frac12}{2\pi\hbar}\left(E+\frac12E_0\right)\ ,
\nonumber\\
\quad\mbox{\rm and}\quad \rho_0^{(3)}\! &=&
 \frac{16}{5}\frac{(E-E_0)^\frac12}{2\pi\hbar}\left(E^2+\frac12EE_0+\frac38E_0^2\right)\ .
 \end{eqnarray}
for the cases of $k=1,2,3$. A general formula, derived in ref.~\cite{Johnson:2023ofr},  is:
\begin{equation}
  \rho_0^{(k)}= \frac{2k}{2^{2k-2}}\sum_{i=1}^k\frac{(2k-1)!(-1)^{i-1}}{(k-i)!(k+i-1)!}\frac{\cos\left[(2i-1)\theta_0\right]}{(2i-1)}\frac{1}{2\pi\hbar}E^{k-\frac12}\quad {\rm with}\quad \sin\theta_0\equiv\sqrt{\frac{E_0}{E}} \ .
\end{equation} Each such $\rho^{(k)}_0$ (treated as an expansion in $E_0/E$) will  be multiplied by  the ${\tilde t}_{k}(E_0)$ contributions determined in step {\bf (1)} to potentially give $B_k(E_0)$.
 This turns out to be not enough to get a match, and so there must be another source of contributions to the $B_k(E_0)$. But there are no more options involving the $x<0$ structure of the multicritical models.

 It turns out there is a third and closely related fourth possibility~\cite{Johnson:2023ofr}:

\item  The value of $\mu$, the end of the $x$-integration in the $x>0$ region might also depend upon $E_0$. But it must do so in a manner that does not affect  the $B_1$ term, which has no $E_0$ dependence. It will be denoted $\tilde\mu$.

\item The simple $u_0(x){=}0$ behaviour in the $x>0$ region must be inadequate. A natural guess is that it is asymptotically of the form deduced in the discussion after equation~(\ref{eq:leading-special})   (or in exploring the large order Bessel models in section~\ref{sec:bessel-models}):  $u_0(x)={\widetilde\Gamma}^2/x^2={\tilde\mu}^2E_0/x^2$ where the second equality comes from the expectation that at the upper limit on the integration $x={\tilde\mu}$, the value of $u_0$  should be the threshold energy~$E_0$. Such a $u_0(x)$ gives a contribution to the spectral density of the form:
 \begin{equation}
 \label{eq:Neq2-positive-integral}
     \frac{1}{2\pi\hbar}\int_0^{\tilde\mu}\frac{\Theta(E-u_0(x))}{\sqrt{E-u_0(x)}}dx =
     \frac{1}{2\pi\hbar}\int^{E}_{E_0}\frac{{\tilde \mu}\sqrt{E_0}du_0}{2u_0^\frac32\sqrt{E-u_0}}
     =
     \frac{\tilde\mu}{2\pi\hbar}\frac{\sqrt{E-E_0}}{E}\ ,
 \end{equation}
 where ${\tilde\mu}=1/2\pi +\cdots$ where the ellipsis indicates $E_0$ dependence. 

\end{enumerate}

 These last two possibilities (expanding in $E_0/E$) allow for new contributions to the $B_k(E_0)$ over and above sources in~{\bf (1)} and~{\bf (2)}. Two miracles happen at this point~\cite{Johnson:2023ofr}. The first is that at any given order in the $E_0/E$ expansion, terms in the expansion of ${\tilde\mu}(E_0)$ in powers of  $E_0$ can be determined that account for the problematic terms in each $B_k(E_0)$. The first few orders determined in this manner are:
 \begin{equation}
     \label{eq:first-few}
{\tilde\mu}(E_0)=\frac{1}{2\pi}\left(1-\frac23\pi^2E_0+\frac{2}{15}\pi^4E_0^2-\frac{4}{315}\pi^6E_0^3+\cdots\right)
 \end{equation}
 
 The second miracle is that the  coefficients of this expansion of ${\tilde\mu}(E_0)$ are such that it can be resummed into a simple expression, and happily the same happens for the $E_0$ expansion of the ${\tilde t}_k$:
\begin{equation}
    \label{eq:mu-awesome}
    {\tilde\mu}(E_0)=\frac{\sin(2\pi\sqrt{E_0})}{4\pi^2\sqrt{E_0}}\ ;\qquad {\tilde t}_k(E_0) = t_k \frac{2^k k!}{(2\pi\sqrt{E_0})^k}J_k(2\pi\sqrt{E_0})\ ,
\end{equation}
where $t_k$ are in equation~(\ref{eq:teekay2}).
This is remarkable, because  from step {\bf (4)} above we have~$\Gamma{=}\pm\hbar^{-1}{\tilde\mu}\sqrt{E_0}$, and using~(\ref{eq:spectral-density-Neq2}) to rewrite~$E_0$ in terms of the charges, it transpires~\cite{Johnson:2023ofr} that (taking the positive root) the number of zero energy states is $\Gamma{=}{\rm e}^{S_0}\sin(\pi q/{\hat q})/4\pi^2$, which precisely agrees with the ${\cal N}{=}2$ BPS formula~(\ref{eq:BPS-density-Neq2})!

 It is worth remarking upon what has emerged. Asking that the spectral density formula~(\ref{eq:spectral-density-Neq2}) be able to be constructed as a matrix model by combining multicritical models resulted in a very specific combination of ingredients that {\it only works} if the precise non-trivial BPS formula~(\ref{eq:BPS-density-Neq2}) is obeyed, with the precise gap of $E_0$. This is very strong evidence of the tight structural connection  between the BPS sector and the gap in the matrix model framework that is enforced by the equation~(\ref{eq:big-string-equation}).

Recall the condition (mentioned just below equation~(\ref{eq:BPS-density-Neq2})) that when $q{>}{\hat q}$, there are no BPS states. Instead of a gap, the spectrum simply begins at $E_0(q)$. This is also accommodated in the formalism.  
As noted in subsection~\ref{sec:degenerate_or_not},  this  is simply  the case of studying~$M{=}QQ^\dagger$  (instead of $Q^\dagger Q$), where the matrix $M$ does not have $\Gamma$ zeros.

\subsection{$\mathcal{N}=4$ JT Supergravity}
\label{sec:Neq4-JT-supergravity}

Emboldened by the success of the ${\cal N}{=}2$ case of the previous section, it is natural to wonder if something similar happens for  ${\cal N}{=}4$ JT supergravity. Such a gravity theory naturally arises~\cite{Heydeman:2020hhw} in the study of near-BPS charged black holes in four dimensional ${\cal N}{=}2$ supergravity, where the near horizon geometry factorizes into AdS$_2\times S^2$. This is reviewed in Appendix~\ref{sec:gravity-embedding}.

In this case we have R-charge $SU(2)_{\rm R}$ with multiplets labelled by the highest spin $J$ representations of this group. After solving the Schwarzian theory, the leading spectral density that emerges is:\footnote{An earlier version of this manuscript had an extra factor of $\frac12$ for the normalization, following  ref.~\cite{Heydeman:2020hhw}, which  was corrected by ref.~\cite{Turiaci:2023jfa}. This has been corrected here now, as well as all the formulae that follow from it in the remainder of this section. The new normalization is consistent with that in ref.~\cite{Turiaci:2023jfa}.}
\begin{equation}
\label{eq:spectral-density-Neq4}
    \rho_0(E) =  {\rm e}^{S_0}\frac{J\sinh\bigl(2\pi\sqrt{E-J^2}\bigr)}{\pi^2 E^2}\Theta\bigl(E-J^2\bigr)\  ,
\end{equation}
with a contribution to the  BPS density (arising only from $J=0$ states):
\begin{equation}
\label{eq:BPS-Neq4}
   \rho_0^{\rm BPS}(E)= {\rm e}^{S_0}\delta(E)\delta(J)\ .
   \end{equation}
As before, a conventional choice has been made to set an embedding scale $E_{\rm sch}{=}2$  in equation~(\ref{eq:spectral-density-Neq4}).

It is natural to wonder to what extent can such a structure  can arise from a random matrix model. Ref.~\cite{Turiaci:2023jfa} made some initial steps to address this problem, starting with the Schwarzian theory and trying to deduce features of the model. The alternative approach used in this paper  will explicitly start from appropriately chosen {\it non-perturbative} matrix model building blocks and see if  they can be put together in a manner that yields a sensible model with this spectrum. 

Expanding~(\ref{eq:spectral-density-Neq4}) in a manner entirely analogous to the previous case, the density will be of the form~(\ref{eq:expansion2}). The hope is that there is some combination of the multicritical models that will 
Again, it is possible to deduce a $J$-dependent formula for some ${\tilde t}_k(J)$ that will  yield the positive powers of $E$ in the expansion. This can be done by working order by order in powers of $J$. After a few orders a pattern appears and just as in the ${\cal N}{=}2$ case, the results can be re-summed into closed form involving Bessel functions.

In retrospect, the results can be readily deduced given that the powers of $E$ being matched are just  shifted down by one unit compared to the previous case, (since there is an extra power of $E$ in the denominator of~(\ref{eq:spectral-density-Neq4}) compared to (\ref{eq:spectral-density-Neq2})). The upshot is that the ${\tilde t}_k$s here must be closely related to the ${\tilde t}_{k+1}$s of the previous section, and the result is:
\begin{equation}
    {\tilde t}^{{\cal N}=4}_k(J) = 8\pi J \frac{2k+2}{2k+1}{\tilde t}^{{\cal N}=2}_{k+1}(E_0\to J^2) = \frac{ J\, 2^{k+4} \pi^{2 k +2}}{\left(4 k^{2}+8 k +3\right) k !}\frac{J_{k+1}(2\pi J)}{(2\pi J)^{k+1}}\ . 
\end{equation}
There are again an infinite number of $B_k$ coefficients controlling negative powers of $E$ that are not correctly matched by using these ${\tilde t}_k(J)$, and so it must be left to the ansatz for the solution in the $x>0$ regime, as in step {\bf (4)} of the previous section. The expectation is that asymptotically  $u_0(x){=}{\widetilde\Gamma}^2/x^2{=}{\tilde\mu}^2J^2/x^2$ for $x>0$, so that when $x={\tilde\mu}$, we get $u_0({\tilde\mu})=J^2$, marking the beginning of the continuum part of the spectrum. Here, ${\tilde\mu}$ is some function of $J$ to be determined. Note that there is a big danger here, since whatever function the procedure gives needs to be such that $|{\widetilde \Gamma}|$ is a constant, consistent with the form of the (BPS) part of the spectrum given in equation~(\ref{eq:BPS-Neq4})! (This is in contrast to the ${\cal N}{=}2$ case~(\ref{eq:BPS-density-Neq2}) where ${\widetilde\Gamma}$ is a non-trivial function of the threshold energy $E_0$, {\it i.e.,}~$q$.)

Undaunted by this, we examine now the negative powers of $E$, seeking conditions on the $J$ dependence of ${\tilde\mu}$. As before, there are terms coming from the subleading powers in the multicritical models (with ${\tilde t}_k$ given above) combined with terms coming from expression~(\ref{eq:Neq2-positive-integral}) now with $E_0\to J^2$.
Working in an expansion of powers of $J$, (and using the ${\tilde t}_k(J)$ already determined) the requirement that the term $B_1/\sqrt{E}$ is reproduced by the construction gives:
\begin{equation}
    \label{eq:mu-expansion}
    {\tilde \mu}=\frac{8 \pi^{2}}{3} J -\frac{16}{15} \pi^{4} J^{3}+\frac{16}{105} \pi^{6} J^{5}-\frac{32}{2835} \pi^{8} J^{7}+\frac{16}{31185} \pi^{10} J^{9}-\frac{32}{2027025} \pi^{12} J^{11}+\mathrm{O}\! \left(J^{13}\right)\ ,
\end{equation}
After some thought, it turns out that this is the series of expansion of a remarkable exact formula:
\begin{equation}
    \label{mu-exact-Neq4}
    {\tilde \mu}= 
    \frac{1}{J}\left(\frac{\sin(2\pi J)}{\pi J}- 2\cos(2\pi J)\right)      
    \ ,
\end{equation}
to which we will return shortly. 
At this point, there are no additional parameters to match,  but there remain contributions to an infinite number of coefficients,~$B_k$ ($k=2,\hdots,\infty$) that simply cannot be reproduced by the matrix model construction and so {\it they must vanish}. This looks hopeless until it becomes clear that they can  be re-summed into simpler expressions that turn out to  all be proportional $\sin(2\pi J)$ (multiplied by the appropriate powers of $J$ needed by dimensional analysis). This is a remarkable outcome, since all these terms  therefore vanish for $J$ integer and half integer which are {\it precisely}  the allowed spin representations of the $SU(2)_{\rm R}$! We have thus succeeded in showing that a spectral density of the precise form~(\ref{eq:spectral-density-Neq4}) can be realized as a matrix model by building it out of multicritical models.
Turning to what this  means for $\widetilde\Gamma$, we see that for the allowed $J$, equation~(\ref{mu-exact-Neq4}) gives ${\tilde\mu}=2/J$, and therefore: 
\begin{equation}
{\widetilde\Gamma}=\pm{\tilde\mu}J=\pm\left(
\frac{\sin(2\pi J)}{\pi J}- 2\cos(2\pi J)   \right)
=\pm2\,, \qquad J \in \frac{1}{2}\mathbb{Z}_{\ne 0}\ .
\end{equation}
 In other words, we  recover (after taking the positive sign, depending upon $J$'s value) that indeed there is a simple dependence $\Gamma={\rm e}^{S_0}\times\text{constant}$, as seen from the Schwarzian analysis of the form of the BPS sector!  
 Crucially, this works {\it only} when $J$ is an allowed value to give the correct the $SU(2)_{\rm R}$ representation. The value of the constant is $2$ here, and not unity as might be expected from~(\ref{eq:BPS-Neq4}). This is because of how states are counted in refs.~\cite{Heydeman:2020hhw,Turiaci:2023jfa},  where the spectrum~(\ref{eq:spectral-density-Neq4}) and~(\ref{eq:BPS-Neq4}) were computed. For non-zero $J$, states of positive and negative $J$ are counted separately, resulting in the doubling. This is different for $J{=}0$, the case for which there is truly a BPS sector.

 Note that putting $J{=}0$ into the above analysis  collapses the predictive structure somewhat since the entire continuum part vanishes altogether, meaning that all $t_k=0$. This still results in a non-trivial leading solution for the string equation~(\ref{eq:big-string-equation}): $u_0={\widetilde\Gamma}^2/x^2$ (discussed in subsection~\ref{sec:bessel-models}), but there is no information from the non-zero $J$ sector to insert. 
 That the $J{=}0$ case is BPS is not something the matrix model might be expected to ``know" explicitly, at least perturbatively: The analysis simply gives that there are exactly ${\rm e}^{S_0}$ states at $E=0$, as long as  the charge $J$ on the continuum multiplet is correct, and with {\it precisely} a gap until energy $E{=}J^2$.  (In followup work on this model, and the one of the last section, presented in ref.~\cite{Johnson:2025oty}, further features are analyzed, including implications from non-perturbative considerations, and we refer the reader there.)
 

 This, our final matrix model exhibit, is  further strong evidence of the tight connection between matrix model ground state degeneracy and the nature of the gap. 
Overall it is remarkable that (in both the ${\cal N}{=}2$ and ${\cal N}{=}4$ cases) the success of a multicritical matrix model construction of the gravity model's non-BPS sector {\it requires} the correct R-charges for the spectrum {\it and} demands the precise number/form of the degenerate ground state sector (BPS) to be present in the model as well. It is clearly of interest to learn more about how (and why) this works, and to see to what else such multicritical model constructions can teach us about JT supergravity and perhaps black holes more generally. See ref.~\cite{Johnson:2025oty} for recent work on this.

\section{Closing Remarks}
\label{sec:discussion}

This paper has focused on the  mechanism within random matrix models that dynamically relates the presence of a large degeneracy with a large spectral gap. It arises from an extreme form of eigenvalue repulsion, the same kind of repulsion that governs other characteristic features of random matrix spectra, giving rise to connections to quantum chaotic systems~\cite{Bohigas:1983er}, and  in recent times to black holes~\cite{Maldacena:2015waa,Cotler:2016fpe,Saad:2019lba}.

Our main point is as follows: Because the low-energy near-horizon geometry of extremal black holes  in $D>2$ factorizes into pieces including  an effective 2D gravity theory, and because the gravitational path integral  for such 2D gravity models is equivalent (order by order in topology) to a random matrix model, the degeneracy-gap connection becomes a {\it prediction} for such black holes, giving an alternative explanation for the presence of the gap that {\it must} appear in order to resolve problematic features of the semi-classical physics of extremal black holes (reviewed in the Introduction).

The overarching lesson is that very general features from the random matrix perspective arise from very delicate gravitational path integral calculations such as those done in refs.~\cite{Heydeman:2020hhw,Boruch:2022tno}. To be sure, we are {\it not} claiming that the gravitational approach is somehow rendered unnecessary by  random matrix model insights. The latter make no reference to crucially important  scales that come from embedding the Schwarzian problem into the gravitational physics. So while the matrix model can predict non-trivially that a gap should appear in the semi-classical/gravity approach, it seems that it cannot on its own determine the scale of the gap. 

Nevertheless, the framework is very compelling. In the precise examples of ${\cal N}{=}2$ and ${\cal N}{=}4$ JT supergravity, arising in the throats of certain important $D{=}4$ and $D{=}5$ supergravity settings (where the extremal black holes are stable, unavoidable solutions) their construction in terms of multicritical models (fully non-perturbatively defined by string equation~(\ref{eq:big-string-equation})) works {\it only} if the precise relation between the BPS formula and the location of the gap is obeyed, as well as requiring the constraints from R-symmetry to be preserved. The lack of freedom to adjust the construction is striking.

There are several open questions that are worth pursuing. Among them are: 

$\bullet$ In this work we have analyzed systems where there are $\mathcal{O}({\rm e}^{S_0})$ degenerate ground states and the scale of the gap in the spectrum 
goes as $E_{\t{schw}} \sim \mathcal{O}(1/S_0)$. The latter is an output of a gravitational calculation, and cannot be deduced from the matrix model itself. It would be interesting if there was a simple explanation of this scaling that also did not rely on a delicate gravity computation. Alternatively, it would be interesting if there were examples where the supergravity gap has a different scaling behavior.


$\bullet$ It would be interesting if there are examples of chaotic systems with a large ground state degeneracy that nonetheless is much smaller than the number of black hole ground states. From random matrix theory expectations one expects a gap to exist whenever there are a large number of degenerate states. Hints of such a gap were found in studies of $\frac{1}{16}$-BPS states in the weak coupling regime of $\mathcal{N}{=}4$ supersymmetric Yang-Mills with small gauge group rank \cite{Chang:2022mjp,Chang:2023zqk}. Surprisingly, there is numerical evidence that in a sector with only one BPS ground state in $SU(N=4)$ super Yang-Mills there is a gap \cite{Chang:2023zqk}, where large $N$ and strong coupling expectations would imply $\sim {\rm e}^{N^2}$ ground states and a gap of order $\sim 1/N^2$. It would be interesting to understand such systems further, see \cite{Chen:2024oqv} for recent work on chaos in the BPS sector. 

$\bullet$ Of course, there is still the matter of studying the perturbative and non-perturbative corrections to the leading order phenomena seen in this paper. The good news is that it is all elegantly captured in the full solution of non-linear ODE that is the string equation~(\ref{eq:big-string-equation}). The leading order asymptotics studied here act as boundary conditions for a smooth solution for $u(x)$, as well as seeds for a recursive asymptotic expansion for $u(x)$ that can be used (following techniques developed recently in refs.\cite{Johnson:2024fkm,Lowenstein:2024gvz}) to define important quantities such as  Weil-Petersson volumes, {\it etc}. (Recent work exploring  this has been done in ref.~\cite{Ahmed:2025lxe}.) It is possible to anticipate some non-perturbative features, based on experience with other cases. Chief among them are small contributions to $\rho(E)$ in the gap region. These will be highly suppressed, of $O({\rm e}^{-{\rm e}^{S_0}})$, but nevertheless non-zero. Other non-perturbative features worth exploring are the underlying microstate probability peaks, as seen in refs.~\cite{Johnson:2021zuo,Johnson:2022wsr}.  In general, however, in contrast to the cases first explored in ref.~\cite{Johnson:2020exp} the string equation is even more  difficult to solve numerically in the regime of interest (large~$\Gamma$), and so there is much work to be done. Since this paper first appeared, follow up work refining the analysis of the string equation for systems with extended supergravity, as well as analyzing aspects of the non-perturbative physics, has appeared in ref.~\cite{Johnson:2025oty}.

\section*{Acknowledgements}

We thank Luca Iliesiu, Maciej Konalowski, Henry Maxfield, and Joaquin Turiaci for helpful discussions and comments. M.U. was supported in part by grant NSF PHY-2309135 to the Kavli Institute for Theoretical Physics (KITP), and by a grant from the Simons Foundation (Grant Number 994312, DG). CVJ's work on this project was supported by US Department of Energy grant  \#DE-SC 0011687, as well as by National Science Foundation grant PHY-2210452  at the Aspen Center for Physics workshop ``The Microscopic Origin of Black Hole Entropy'', and National Science Foundation grant PHY-1748958 at the KITP workshop ``What is String Theory?''. CVJ also thanks Amelia for her support and patience.


\bigskip
\bigskip

\appendix

\noindent 
{\bf \Large Appendix}

\section{Gravitational interpretation}
\label{sec:gravity-embedding}
In the body of the paper  we considered  matrix model analyses for the emergence of a gap in the spectrum when there is a large ground state degeneracy. In this appendix we will briefly review gravitational systems described by these matrix models, and how the features that naturally emerge from the random matrix analysis emerge from a gravitational description.
The general procedure to study the spectrum of black holes in a given gravitational theory is to compute a thermal partition function using the gravitational path integral:
\be
Z(\beta, \alpha_i) = \Tr \lr{ {\rm e}^{-\beta H + i \alpha_i J_i} }\,,
\ee
where in the above the specification of $\beta$ and chemical potentials $\alpha_i$ fixes the asymptotic boundary conditions for our path integral. The periodicity of the thermal circle is set to $\beta$, and we have allowed for some generic set of conserved charges $J_i$ that depend on the details of the theory with $\alpha_i$ setting theory dependent boundary conditions. Evaluating the above expression by summing over gravitational configurations, we can extract the density of states at fixed values of the charges by performing an inverse Laplace transform
\be
\rho(E, j_i) = \int_{i \mathbb{R}} d \beta \int_{\mathbb{R}} d \alpha_i ~ {\rm e}^{\beta E -i\alpha_i j_i} Z(\beta, \alpha_i)\,.
\ee
The evaluation of the thermal partition function amounts to summing over all Euclidean saddle points that satisfy the asymptotic boundary conditions, and including loop effects around each saddle. For a variety of near-extremal black holes in diverse spacetime dimensions and theories, as the temperature is lowered to zero $\beta \to \infty$, the path integral simplifies and the dynamics reduce to an effective Schwarzian mode which should be thought of as living in the throat of the near-extremal black hole being studied \cite{Iliesiu:2020qvm,Heydeman:2020hhw,Boruch:2022tno,Iliesiu:2022onk,Turiaci:2023wrh}. For non-supersymmetric black holes without a large ground state degeneracy the dynamics are given by a non-supersymmetric Schwarzian \cite{Iliesiu:2020qvm}.

We are interested in the case of supersymmetric black holes with a large number of ground states. In this case the effective dynamics close enough to extremality are given by an $\mathcal{N}=2$ or $\mathcal{N}=4$ super Schwarzian theory \cite{Heydeman:2020hhw}. We discuss the case of $\mathcal{N}=2$, which has action \cite{Fu:2016vas}:
\be
I_{\mathcal{N}=2}= \frac{2\pi}{\beta \eschw } \int_0^{2\pi} d \tau \lr{-\operatorname{Sch}(f, \tau)+2\left(\partial_\tau \sigma\right)^2 } + (\t{fermionic action}) \,.
\ee
In the above $f, \sigma$ are bosonic fields defined on the circle with $\sigma$ being periodically identified $\sigma \sim \sigma + 2\pi n \hat{q}$, while the fermionic action depends on anti-periodic fermions $\eta, \bar \eta$. Turning on the R-charge imposes the boundary conditions $\sigma(\tau + 2\pi) = \sigma (\tau) + \hat{q} \alpha, ~\eta(\tau + 2\pi) = - {\rm e}^{i \hat{q} \alpha} \eta(\tau), ~ \bar \eta(\tau + 2\pi) = -{\rm e}^{i \hat{q} \alpha} \bar \eta(\tau)$.\footnote{We work in conventions where we are computing $\Tr ({\rm e}^{-\beta H + i \alpha Q})$ and $Q$ has R-charge $\hat{q}$.} The parameters $\hat{q}$ and $\eschw$ come from the dimensional reduction around the black hole background being studied. The partition function is given by \cite{Stanford:2017thb,Mertens:2017mtv,Turiaci:2023jfa}:
\be
Z(\beta, \alpha) = \sum_{n \in \mathbb{Z}} \frac{\hat{q}}{2\pi^3} \frac{\cos \lr{\pi \hat{q} (n+\frac{\alpha}{2\pi})}}{1-4 \hat{q}^2 (n+\frac{\alpha}{2\pi})^2 } \exp \left[S_0 + \frac{2\pi^2}{\beta \eschw} \lr{1-4 \hat{q}^2 \lr{n+\frac{\alpha}{2\pi}}^2} \right]\,.
\ee
The sum over classical saddles $f=\tan \frac{\tau}{2}, \sigma = \hat{q}(n+\frac{\alpha}{2\pi})\tau$ is given by the sum over $n$, with only bosonic fields contributing to the classical action. Including bosonic and fermionic one-loop determinants removing appropriate zero modes gives a temperature independent prefactor to the exponential. 

We first comment on the appearance of BPS states which have zero energy. The presence of BPS states can be diagnosed by taking the zero temperature limit $\beta \to \infty$ and seeing if the partition function vanishes. From the above,
\be
\lim_{\beta \to \infty} Z(\beta,\alpha) = {\rm e}^{S_0} \sum_{n \in \mathbb{Z}} \frac{\hat{q}}{2\pi^3}\frac{\cos \lr{\pi \hat{q} (n+\frac{\alpha}{2\pi})}}{1-4 \hat{q}^2 (n+\frac{\alpha}{2\pi})^2 } \,,
\ee
giving a large ground state degeneracy of order ${\rm e}^{S_0}$. Note that the contribution of BPS states to the partition function crucially relies on the one-loop determinants. Namely, the temperature dependence in the bosonic and fermionic determinants cancel each other out, giving a temperature independent prefactor. The origin of this cancellation is that there are an equal number of bosonic and fermionic zero modes $n^0_b=n^0_f=4$, with the one-loop prefactor scaling as $Z_{\t{1-loop}} \propto \beta^{n^0_f-n^0_b}$. 

Since there is a large degeneracy we expect the appearance of a gap. If we naively take the inverse Laplace transform of a single saddle-point we would find a density of states:
\be
\rho_n(E) = {\rm e}^{S_0} \frac{\cos \lr{\pi \hat{q} (n+\frac{\alpha}{2\pi})}}{1-4 \hat{q}^2 (n+\frac{\alpha}{2\pi})^2 } \lr{\delta(E) + \sqrt{\frac{c_n}{E}} I_1 \lr{\sqrt{4 c_n E}} }, \quad c_n = \frac{2 \pi^2}{\eschw} \lr{1-4 \hat{q}^2 \lr{n+\frac{\alpha}{2\pi}}^2}\,,
\ee
which has support down to zero energy and becomes negative for certain segments of energies that depends on the saddle index $n$. To see that there is a gap and to extract the spectrum it is most convenient to perform a Poisson re-summation and re-express the sum over saddles instead as a sum over R-charges, giving \cite{Mertens:2017mtv,Turiaci:2023jfa}:
\begin{gather}
Z(\beta,\alpha) = \sum_{k\in \mathbb{Z}, |k|< \frac{\hat{q}}{2}} {\rm e}^{i \alpha k} N_{\t{BPS}}(k) + \sum_{k \in \mathbb{Z}}\lr{{\rm e}^{i \alpha k } +{\rm e}^{i \alpha (k+\hat{q})}}\int_{E_0(k)}^\infty d E \rho_k (E) {\rm e}^{-\beta E}\,,\\
\rho_k(E) =  \frac{{\rm e}^{S_0}\sinh{\lr{2\pi\sqrt{2}\sqrt{ \frac{E-E_0(k)}{\eschw}}}}}{8\pi^3 E}, \qquad E_0(k) = \frac{\eschw}{8} \lr{\frac{k}{\hat{q}}+\frac{1}{2}}^2 \,, \\
N_{\t{BPS}}(k) = \frac{{\rm e}^{S_0} \cos \frac{\pi k}{\hat{q}}}{4\pi^2}\,, \qquad \t{for} \qquad |k| < \frac{\hat{q}}{2}\,.
\end{gather}
In the above the BPS states have R-charges $|k| < \frac{\hat{q}}{2}$ with a degeneracy of order ${\rm e}^{S_0}$. Non-BPS states fall into multiplets of charges $(k, k + \hat{q})$ with density $\rho_k(E)$, and they exhibit a gap from the BPS states given by $E_0(k)$ for $|k| < \frac{\hat{q}}{2}$. For $|k| > \frac{\hat{q}}{2}$ there are no BPS states at zero energy and the spectrum starts at some finite value $E_0(k)>0$.\footnote{There exist variants of the $\mathcal{N}=2$ model that have an anomaly where the allowed values of the R-charge are shifted. In such models it's possible to have sectors where the continuum states begin at zero energy $E_0(-\frac{\hat{q}}{2})=0$, but in such cases the number of BPS states is zero $N_{\t{BPS}}(-\frac{\hat{q}}{2})=0$ and so there is no contradiction with a large number of states generating a gap. See \cite{Turiaci:2023jfa} for additional discussion.} These formulae precisely match the the matrix model results in section \ref{sec:Neq2-JT-supergravity}, with $k$ replaced by $q-\frac{{\hat q}}{2}$.

\paragraph{Gravitational system with $\mathcal{N}{=}2$ Schwarzian.} An example gravitational system where the $\mathcal{N}=2$ Schwarzian appears is for BPS black holes dual to $1/16$-BPS states in $\mathcal{N}=4$ super Yang-Mills and their corresponding near-BPS black hole cousins \cite{Boruch:2022tno}. From the AdS$_5$ perspective these are rotating, charged black holes. At extremality the near-horizon geometry of these black holes is $AdS_2 \times S^3 \times S^5$ and the Schwarzian lives in the $AdS_2$ throat. The parameters of the effective Schwarzian theory are $\hat{q}=1, S_0 \propto N^2, \eschw \propto \frac{1}{N^2}$. The pre-factors are complicated $\mathcal{O}(1)$ constants, and we have only given their scaling with $N$ of the boundary gauge group $U(N)$. The extremal BPS entropy of these black holes is $S_0$, and $\eschw$ is an emergent energy scale that sets the size of the gap in the spectrum between BPS and near-BPS states. 

For such black holes we have a large ground state degeneracy ${\rm e}^{S_0}$ at charge $q=0$ with a gap in the spectrum to the nearest near-BPS BHs starting at $E_0 = \eschw/32$. For other charge sectors there are no BPS states, and the spectrum begins at the value $E_0(q)$ quoted above. How the gap scale $\eschw$ scales with the extremal entropy $S_0$ cannot be determined from the two dimensional theory, the precise scaling arises from the details of the dimensional reduction. 

\paragraph{Gravitational system with $\mathcal{N}{=}4$ Schwarzian.} Gravitational systems with an emergent $\mathcal{N}=4$ Schwarzian description were partially studied in \cite{Heydeman:2020hhw}. A very simple example are four dimensional near-BPS near-extremal charged black holes with AdS$_2 \times S^2$ near horizon throats. These black holes arise in $\mathcal{N}=2$ supergravity, and their near-horizon dynamics are governed by the action \cite{Heydeman:2020hhw}:
\be
I_{\mathcal{N}=4}= -\frac{2\pi}{\beta \eschw } \int_0^{2\pi} d \tau \lr{\operatorname{Sch}(f, \tau)+ \Tr \left( g^{-1}\partial_\tau g\right)^2 } + (\t{fermionic action}) \,,
\ee
where now the field $g$ is valued in $SU(2)$, and the theory has an $SU(2)$ R-symmetry. The corresponding charge is identified with the angular momentum on $S^2$ of the near-extremal black hole. The partition function can be evaluated and the density of states is given by \cite{Heydeman:2020hhw,Turiaci:2023jfa}:
\be
\rho_{\mathbf{J}}(E) = {\rm e}^{S_0} \delta(E)\delta(\mathbf{J}) + \frac{{\rm e}^{S_0} \mathbf{J} \eschw }{2\pi^2  E^2} \sinh \lr{ 2 \pi \sqrt{\frac{2 (E-E_0(\mathbf{J}))}{\eschw}}} \Theta(E-E_0(\mathbf{J}))\,, \quad E_0(\mathbf{J}) = \frac{\mathbf{J}^2}{2} \eschw\,.
\ee
States are organized into supermultiplets labeled by the largest angular momentum $\mathbf{J} {=} J \oplus 2 (J{-}\frac{1}{2}) \oplus J{-}1$. The supermultiplet with highest spin $\mathbf{J}$ has density of states given above. We see that the ground state is purely populated by the reduced $\mathbf{J}=0$ multiplet. There exists a continuum of $J{=}0$ states above the ground state separated by gaps $E_0(\mathbf{\frac{1}{2}})$ and $E_0(\mathbf{1})$ since the $\mathbf{1, \frac{1}{2}}$ multiplets contain~$J{=}0$.

The parameters of the $\mathcal{N}{=}4$ Schwarzian theory can again be deduced by dimensional reduction around the extremal Reissner-Nordstrom background. Choosing boundary conditions to fix the asymptotic electric charge in flat space to $Q$, one obtains $S_0 = \pi Q^2$ and $\eschw=1/Q^2$ in units of the AdS$_2$ length. We see that for such black holes the gap between the BPS states and continuum states is given by $E_0(\mathbf{\frac{1}{2}})=1/(8 Q^2)$.

\bibliographystyle{utphys}
\bibliography{Refs.bib,super_JT_gravity1,super_JT_gravity2}
\end{document}